\documentclass{elsart_sf}
\usepackage{epsfig}
\usepackage{amssymb, amsmath}

\begin{document}

\hyphenation{Dépar-te-ment}



%
\begin{center}
\title{Time Calibration of the ANTARES Neutrino Telescope}
\end{center}

\begin{center}
{\bf The ANTARES Collaboration}\\

\vspace{1mm}
\author[IFIC]{J.A. Aguilar},
\author[CPPM]{I. Al Samarai},
\author[Colmar]{A. Albert},
\author[Barcelona]{M. Andr\'{e}},
\author[Genova]{M. Anghinolfi},
\author[Erlangen]{G. Anton},
\author[IRFU/SEDI]{S. Anvar},
\author[UPV]{M. Ardid},
\author[NIKHEF]{A.C. Assis Jesus},
\author[NIKHEF]{T.~Astraatmadja}\footnote{Also at University of Leiden, the Netherlands},
\author[CPPM]{J.J. Aubert},
\author[Erlangen]{R. Auer},
\author[APC]{B. Baret},
\author[LAM]{S. Basa},
\author[Bologna-UNI,Bologna]{M. Bazzotti},
\author[CPPM]{V. Bertin},
\author[Bologna-UNI,Bologna]{S. Biagi},
\author[IFIC]{C. Bigongiari},
\author[UPV]{M. Bou-Cabo},
\author[NIKHEF]{M.C. Bouwhuis},
\author[CPPM]{A. M. Brown},
\author[CPPM]{J.~Brunner}\footnote{On leave at DESY, Platanenallee 6, D-15738 Zeuthen, Germany},
\author[CPPM]{J. Busto},
\author[UPV]{F. Camarena},
\author[Roma-UNI,Rome]{A. Capone},
\author[Clermont-Ferrand]{C.C$\mathrm{\hat{a}}$rloganu},
\author[Bologna-UNI,Bologna]{G. Carminati},
\author[CPPM]{J. Carr},
\author[Bologna,INAF]{S. Cecchini},
\author[GEOAZUR]{Ph. Charvis},
\author[Bologna]{T. Chiarusi},
\author[Bari]{M. Circella},
% ...
\footnote{Corresponding author: zornoza@ific.uv.es},
\author[IFIC]{J.~Z\'u\~{n}iga}

\thanks[tag:1]{\scriptsize{Also at University of Leiden, the Netherlands}}
\thanks[tag:2]{\scriptsize{On leave at DESY, Platanenallee 6, 15738 Zeuthen, Germany}}

\newpage
\nopagebreak[3]

\address[IFIC]{\scriptsize{IFIC - Instituto de F\'isica Corpuscular, Edificios Investigaci\'on de Paterna, CSIC - Universitat de Val\`encia, Apdo. de Correos 22085, 46071 Valencia, Spain}}\vspace*{0.10cm}
\nopagebreak[3]
\vspace*{-0.20\baselineskip}
\nopagebreak[3]
\address[CPPM]{\scriptsize{CPPM - Centre de Physique des Particules de Marseille, CNRS/IN2P3 et Universit\'e de la M\'editerran\'ee, 163 Avenue de Luminy, Case 902, 13288 Marseille Cedex 9, France}}\vspace*{0.10cm}
\nopagebreak[3]
\vspace*{-0.20\baselineskip}
\nopagebreak[3]
\address[Colmar]{\scriptsize{GRPHE - Institut universitaire de technologie de Colmar, 34 rue du Grillenbreit BP 50568 - 68008 Colmar, France }}\vspace*{0.10cm}
\nopagebreak[3]
\vspace*{-0.20\baselineskip}
\nopagebreak[3]
\address[Barcelona]{\scriptsize{Technical University of Catalonia,Laboratory of Applied Bioacoustics,Rambla Exposició,08800 Vilanova i la Geltrú,Barcelona, Spain}}\vspace*{0.10cm}
\nopagebreak[3]
\vspace*{-0.20\baselineskip}
\nopagebreak[3]
\address[Genova]{\scriptsize{INFN - Sezione di Genova, Via Dodecaneso 33, 16146 Genova, Italy}}\vspace*{0.10cm}
\nopagebreak[3]
\vspace*{-0.20\baselineskip}
\nopagebreak[3]
\address[Erlangen]{\scriptsize{Friedrich-Alexander-Universit\"{a}t Erlangen-N\"{u}rnberg, Erlangen Centre for Astroparticle Physics, Erwin-Rommel-Str. 1, 91058 Erlangen, Germany}}\vspace*{0.10cm}
\nopagebreak[3]
\vspace*{-0.20\baselineskip}
\nopagebreak[3]
\address[IRFU/SEDI]{\scriptsize{Direction des Sciences de la Mati\`ere - Institut de recherche sur les lois fondamentales de l'Univers - Service d'Electronique des D\'etecteurs et d'Informatique, CEA Saclay, 91191 Gif-sur-Yvette Cedex, France}}\vspace*{0.10cm}
\nopagebreak[3]
\vspace*{-0.20\baselineskip}
\nopagebreak[3]
\address[UPV]{\scriptsize{Institut d'Investigaci\'o per a la Gesti\'o Integrada de Zones Costaneres (IGIC) - Universitat Polit\`ecnica de Val\`encia. C/  Paranimf 1, 46730 Gandia, Spain.}}\vspace*{0.10cm}
\nopagebreak[3]
\vspace*{-0.20\baselineskip}
\nopagebreak[3]
\address[NIKHEF]{\scriptsize{FOM Instituut voor Subatomaire Fysica Nikhef, Science Park 105, 1098 XG Amsterdam, The Netherlands}}\vspace*{0.10cm}
\nopagebreak[3]
\vspace*{-0.20\baselineskip}
\nopagebreak[3]
\address[APC]{\scriptsize{APC - Laboratoire AstroParticule et Cosmologie, UMR 7164 (CNRS, Universit\'e Paris 7 Diderot, CEA, Observatoire de Paris) 10, rue Alice Domon et L\'eonie Duquet 75205 Paris Cedex 13,  France}}\vspace*{0.10cm}
\nopagebreak[3]
\vspace*{-0.20\baselineskip}
\nopagebreak[3]
\address[LAM]{\scriptsize{LAM - Laboratoire d'Astrophysique de Marseille, P\^ole de l'\'Etoile Site de Ch\^ateau-Gombert, rue Fr\'ed\'eric Joliot-Curie 38,  13388 Marseille Cedex 13, France }}\vspace*{0.10cm}
\nopagebreak[3]
\vspace*{-0.20\baselineskip}
\nopagebreak[3]
\address[Bologna-UNI]{\scriptsize{Dipartimento di Fisica dell'Universit\`a, Viale Berti Pichat 6/2, 40127 Bologna, Italy}}\vspace*{0.10cm}
\nopagebreak[3]
\vspace*{-0.20\baselineskip}
\nopagebreak[3]
\address[Bologna]{\scriptsize{INFN - Sezione di Bologna, Viale Berti Pichat 6/2, 40127 Bologna, Italy}}\vspace*{0.10cm}
\nopagebreak[3]
\vspace*{-0.20\baselineskip}
\nopagebreak[3]
\address[Roma-UNI]{\scriptsize{Dipartimento di Fisica dell'Universit\`a La Sapienza, P.le Aldo Moro 2, 00185 Roma, Italy}}\vspace*{0.10cm}
\nopagebreak[3]
\vspace*{-0.20\baselineskip}
\nopagebreak[3]
\address[Rome]{\scriptsize{INFN -Sezione di Roma, P.le Aldo Moro 2, 00185 Roma, Italy}}\vspace*{0.10cm}
\nopagebreak[3]
\vspace*{-0.20\baselineskip}
\nopagebreak[3]
\address[Clermont-Ferrand]{\scriptsize{Laboratoire de Physique Corpusculaire, IN2P3-CNRS, Universit\'e Blaise Pascal, Clermont-Ferrand, France}}\vspace*{0.10cm}
\nopagebreak[3]
\vspace*{-0.20\baselineskip}
\nopagebreak[3]
\address[INAF]{\scriptsize{INAF-IASF, via P. Gobetti 101, 40129 Bologna, Italy}}\vspace*{0.10cm}
\nopagebreak[3]
\vspace*{-0.20\baselineskip}
\nopagebreak[3]
\address[GEOAZUR]{\scriptsize{G\'eoazur - Universit\'e de Nice Sophia-Antipolis, CNRS/INSU, IRD, Observatoire de la C\^ote d'Azur and Universit\'e Pierre et Marie Curie, BP 48, 06235 Villefranche-sur-mer, France}}\vspace*{0.10cm}
\nopagebreak[3]
\vspace*{-0.20\baselineskip}
\nopagebreak[3]
\address[Bari]{\scriptsize{INFN - Sezione di Bari, Via E. Orabona 4, 70126 Bari, Italy}}\vspace*{0.10cm}
\nopagebreak[3]
\vspace*{-0.20\baselineskip}
\nopagebreak[3]
\address[IRFU/SPP]{\scriptsize{Direction des Sciences de la Mati\`ere - Institut de recherche sur les lois fondamentales de l'Univers - Service de Physique des Particules, CEA Saclay, 91191 Gif-sur-Yvette Cedex, France}}\vspace*{0.10cm}
\nopagebreak[3]
\vspace*{-0.20\baselineskip}
\nopagebreak[3]
\address[COM]{\scriptsize{COM - Centre d'Oc\'eanologie de Marseille, CNRS/INSU et Universit\'e de la M\'editerran\'ee, 163 Avenue de Luminy, Case 901, 13288 Marseille Cedex 9, France}}\vspace*{0.10cm}
\nopagebreak[3]
\vspace*{-0.20\baselineskip}
\nopagebreak[3]
\address[LNS]{\scriptsize{INFN - Laboratori Nazionali del Sud (LNS), Via S. Sofia 62, 95123 Catania, Italy}}\vspace*{0.10cm}
\nopagebreak[3]
\vspace*{-0.20\baselineskip}
\nopagebreak[3]
\address[UPS]{\scriptsize{Universit\'e Paris-Sud 11 - D\'epartement de Physique, 91403 Orsay Cedex, France}}\vspace*{0.10cm}
\nopagebreak[3]
\vspace*{-0.20\baselineskip}
\nopagebreak[3]
\address[Pisa-UNI]{\scriptsize{Dipartimento di Fisica dell'Universit\`a, Largo B. Pontecorvo 3, 56127 Pisa, Italy}}\vspace*{0.10cm}
\nopagebreak[3]
\vspace*{-0.20\baselineskip}
\nopagebreak[3]
\address[Pisa]{\scriptsize{INFN - Sezione di Pisa, Largo B. Pontecorvo 3, 56127 Pisa, Italy}}\vspace*{0.10cm}
\nopagebreak[3]
\vspace*{-0.20\baselineskip}
\nopagebreak[3]
\address[IPHC]{\scriptsize{IPHC-Institut Pluridisciplinaire Hubert Curien - Universit\'e de Strasbourg et CNRS/IN2P3  23 rue du Loess, BP 28,  67037 Strasbourg Cedex 2, France}}\vspace*{0.10cm}
\nopagebreak[3]
\vspace*{-0.20\baselineskip}
\nopagebreak[3]
\address[NIOZ]{\scriptsize{Royal Netherlands Institute for Sea Research (NIOZ), Landsdiep 4,1797 SZ 't Horntje (Texel), The Netherlands}}\vspace*{0.10cm}
\nopagebreak[3]
\vspace*{-0.20\baselineskip}
\nopagebreak[3]
\address[Bamberg]{\scriptsize{Dr. Remeis Sternwarte Bamberg, Sternwartstra\ss e 7,Bamberg,Germany}}\vspace*{0.10cm}
\nopagebreak[3]
\vspace*{-0.20\baselineskip}
\nopagebreak[3]
\address[KVI]{\scriptsize{Kernfysisch Versneller Instituut (KVI), University of Groningen, Zernikelaan 25, 9747 AA Groningen, The Netherlands}}\vspace*{0.10cm}
\nopagebreak[3]
\vspace*{-0.20\baselineskip}
\nopagebreak[3]
\address[UU]{\scriptsize{Universiteit Utrecht, Faculteit Betawetenschappen, Princetonplein 5, 3584 CC Utrecht, The Netherlands}}\vspace*{0.10cm}
\nopagebreak[3]
\vspace*{-0.20\baselineskip}
\nopagebreak[3]
\address[UvA]{\scriptsize{Universiteit van Amsterdam, Instituut voor Hoge-Energie Fysika, Science Park 105, 1098 XG Amsterdam, The Netherlands}}\vspace*{0.10cm}
\nopagebreak[3]
\vspace*{-0.20\baselineskip}
\nopagebreak[3]
\address[MSU]{\scriptsize{Moscow State University,Skobeltsyn Institute of Nuclear Physics,Leninskie gory, 119991 Moscow, Russia}}\vspace*{0.10cm}
\nopagebreak[3]
\vspace*{-0.20\baselineskip}
\nopagebreak[3]
\address[Catania-UNI]{\scriptsize{Dipartimento di Fisica ed Astronomia dell'Universit\`a, Viale Andrea Doria 6, 95125 Catania, Italy}}\vspace*{0.10cm}
\nopagebreak[3]
\vspace*{-0.20\baselineskip}
\nopagebreak[3]
\address[WIN]{\scriptsize{University of Wisconsin - Madison, 53715, WI, USA}}\vspace*{0.10cm}
\nopagebreak[3]
\vspace*{-0.20\baselineskip}
\nopagebreak[3]
\address[ISS]{\scriptsize{Institute for Space Sciences, R-77125 Bucharest, M\u{a}gurele, Romania     }}\vspace*{0.10cm}
\nopagebreak[3]
\vspace*{-0.20\baselineskip}
\nopagebreak[3]
\address[Genova-UNI]{\scriptsize{Dipartimento di Fisica dell'Universit\`a, Via Dodecaneso 33, 16146 Genova, Italy}}\vspace*{0.10cm}
\nopagebreak[3]
\vspace*{-0.20\baselineskip}
\nopagebreak[3]
\address[ITEP]{\scriptsize{ITEP - Institute for Theoretical and Experimental Physics, B. Cheremushkinskaya 25, 117218 Moscow, Russia}}\vspace*{0.10cm}
\nopagebreak[3]

\end{center}

\begin{frontmatter}
\begin{abstract}

The ANTARES deep-sea neutrino telescope comprises a three-dimensional
array of photomultipliers to detect the Cherenkov light induced by
upgoing relativistic charged particles originating from neutrino
interactions in the vicinity of the detector. The large scattering
length of light in the deep sea facilitates an angular resolution of a
few tenths of a degree for neutrino energies exceeding 10 TeV. In
order to achieve this optimal performance, the time calibration
procedures should ensure a relative time calibration between the
photomultipliers at the level of $\sim$1ns. The methods developed to
attain this level of precision are described.

\end{abstract}

\end{frontmatter}

\section{Introduction}

The ANTARES Collaboration has constructed a neutrino telescope in the
Mediterranean Sea~\cite{bib:icrc}. The main aim of the project is the
search for high-energy neutrinos of astrophysical origin. This is
achieved by the detection of Cherenkov photons induced by the passage
of relativistic charged particles resulting from neutrino interactions
in the material surrounding the detector, the most important channel
being the charged current interactions producing muons. Other
signatures, such as the cascades produced both in charged and neutral
currents are also detected. These photons are detected in a large
three-dimensional array of 885 photomultipliers (PMTs) installed on
twelve vertical lines anchored on the sea bed at a depth of
2475~m. First physics results have been
published~\cite{bib:line1,bib:dmitry,bib:muon}.

Detailed simulations have shown~\cite{bib:aart} that the theoretically
achievable angular resolution of the telescope (crucial for the search
for cosmic point sources) in the muon channel reaches a few tenths of
a degree at energies larger than 10 TeV. This sub-degree resolution
has been preliminary confirmed on data~\cite{bib:antoine}. The
reconstruction of the tracks is based on the probability density
function (PDF) of arrival times of photons at the photomultipliers. A
good timing calibration is therefore mandatory to ensure the
reliability of the track reconstruction algorithms. The precision
required for the relative time calibration system derives from the
transit time spread (TTS) of the photomultipliers
($\sigma_{TTS}\sim$1.3~ns) and the effect of the chromatic dispersion
in water ($\sim$1.5~ns for a typical light path of 40~m). The addition
in quadrature of these contributions leads to a best possible time
resolution of $\sim$2~ns per single photon.

The measurement of the detection times of the PMT hits is based on
clock signals which are distributed from shore so that a common
reference time is used in the whole apparatus, as explained in
Section~\ref{sec:clock}. The clock distribution system determines the
propagation delays for these signals to reach the different
electronics containers offshore with sub-nanosecond precision. The
time measurements from individual PMTs have however to be corrected by
appropriate offsets, which are determined {\it in situ} by means of a
calibration system based on LED and laser devices (optical beacons)
emitting light flashes at known times~\cite{bib:obpaper}. In order to
ensure that the apparatus reaches its maximum pointing accuracy, it is
necessary that the accuracy with which all such offsets are
determined, in the so-called relative time calibration, remains below
1~ns, so that the resulting systematic error is significantly smaller
than the intrinsic event-by-event fluctuations.

The absolute time calibration is less demanding. A precision of a few
seconds is sufficient to correlate reconstructed neutrino directions
with steady point sources and an accuracy of the order of milliseconds
is sufficient for association to transient astrophysical events (GRBs,
AGN flares, SGR bursts, etc.) As it will be explained later, the GPS
system provides a good enough time stamping.

The structure of this article is as follows. A brief description of
the ANTARES detector and its data-acquisition (DAQ) system is given in
Section~\ref{sec:antares}. In Section~\ref{sec:clock}, the main
characteristics of the master clock of the experiment, used to
synchronize all detector elements, are presented.
Section~\ref{sec:omtimes} describes the calibration of the time
offsets specific to each PMT performed in the laboratory
before deployment and {\it in situ}.  In Section~\ref{sec:timeres}
independent cross-checks of the system performance are presented.

\section{The ANTARES detector}
\label{sec:antares}

\begin{figure}
 \begin{center}
 \epsfxsize=10cm
 \epsffile{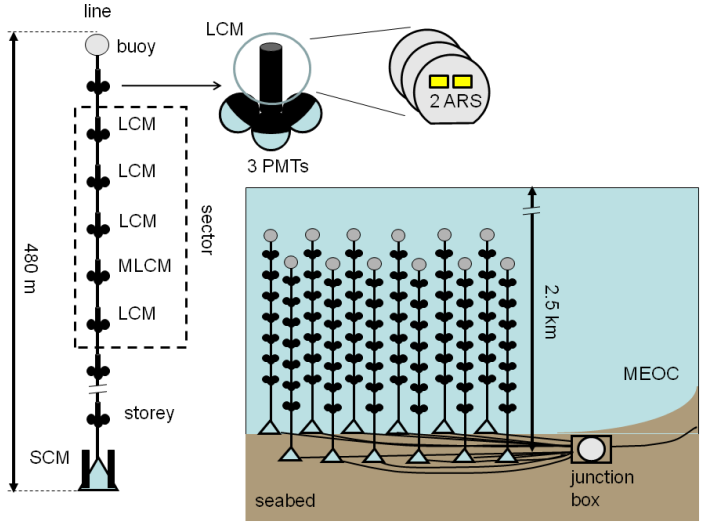}
\caption{\small Schematic view of the ANTARES detector, consisting of
twelve mooring lines connected to the shore station through an
electro-optical cable. Laser beacons are located at the bottom of
Lines 7 and 8 (central lines). LED beacons are located in storeys 2,
9, 15 and 21 (counted from bottom to top) of each line.}
\label{fig:detector}
 \end{center}
\end{figure}

The ANTARES detector is located in the Mediterranean Sea, about 40~km
from Toulon, off the French coast. It consists of 885 10-inch PMTs
(model R7081-20 from Hamamatsu) distributed over twelve lines, of 25
triplet PMT storeys each, anchored to the sea bottom (see
Figure~\ref{fig:detector}). The total length of each line is
480~m. The lowest storey of each line is located 100~m above the sea
bed. The distance between consecutive storeys is 14.5~m and the
horizontal separation between lines is 60\--80~m.  The
PMTs~\cite{bib:pmtpaper} are coupled with optical gel to the inner
surfaces of pressure-resistant glass spheres (optical modules,
OMs~\cite{bib:ompaper}).
A $\mu$-metal cage inside the OM shields the PMT from the Earth
magnetic field. Each storey comprises a triplet of OMs and an
electronics module (Local Control Module, LCM).  Five consecutive
storeys along the line are grouped into a ``sector'' sharing common
data and clock distribution elements.  One LCM of a sector (Master
Local Control Module, MLCM) contains an Ethernet switch and a Dense
Wavelength Division Multiplexing transceiver of a specific wavelength
to establish data connection from the sector to shore along optical
fibres.  At the bottom of each line, a String Control Module (SCM)
houses the power units and the optical (de)multiplexers. Furthermore,
each line incorporates four LED Optical Beacons
(OBs)~\cite{bib:obpaper} used for time calibration and studies of the
optical properties of the sea water. In addition, two Laser Beacons
are located at the bottom of two of the central lines.

The lines are connected to a junction box from where the Main
Electro-Optical Cable (MEOC), of a length of 45~km, links the detector
to the shore station. The shore station houses a computer farm for
data filtering and storage and the master clock. A more detailed
description of the ANTARES DAQ system can be found in~\cite{bib:daq}.

\begin{figure}
 \begin{center}
 \epsfxsize=10cm
 \epsffile{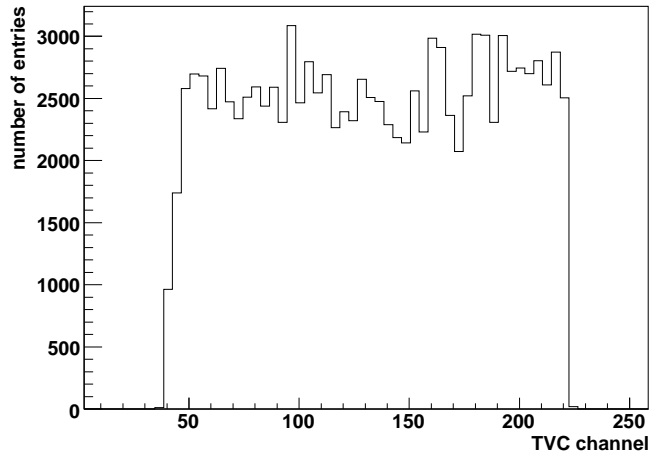}
\caption{\small TVC distribution generated from random PMT
signals. The TVC values have been digitized by an ADC. The histogram
has been rebinned by a factor 4 to mitigate the effect of differential
non-linearities.}
\label{fig:tvc}
 \end{center}
\end{figure}

The signals from one PMT are processed {\it in situ} by ASICs (named
Analogue Ring Samplers or ARSs~\cite{bib:ars}), located in the
LCMs. Two ARSs are used in alternation on each OM in order to reduce
the dead time of the PMT signals close in time. If an analogue PMT
signal reaches an amplitude larger than a given threshold (typically
0.3~photo-electrons), its arrival time and charge are digitized by the
ARS chip. The coarse signal time (time stamp) is given by a counter
which is incremented by a 20 MHz clock synchronized to signals
received from shore by the clock system (see next section). A
Time-to-Voltage Converter (TVC) read by an 8-bit ADC is used to
measure the fine time in the 50 ns period after the time stamp. The
dynamic range of the TVC is determined by recording random signals
uniformly distributed in time (see Figure~\ref{fig:tvc}). The clock
period of 50 ns covers about 150 channels of the ADC, corresponding to
an average bin width of 0.3 ns. The resulting effective time
resolution is thus about 0.3 ns/$\sqrt{12}$. However, differential
non-linearities of the ADCs (i.e. not all bins have the same width)
result in an effective precision about four times larger, which is
still adequate for our requirements to keep time calibration
resolution below $\sim$1~ns.

\begin{figure}
 \begin{center}
 \epsfxsize=10cm
 \epsffile{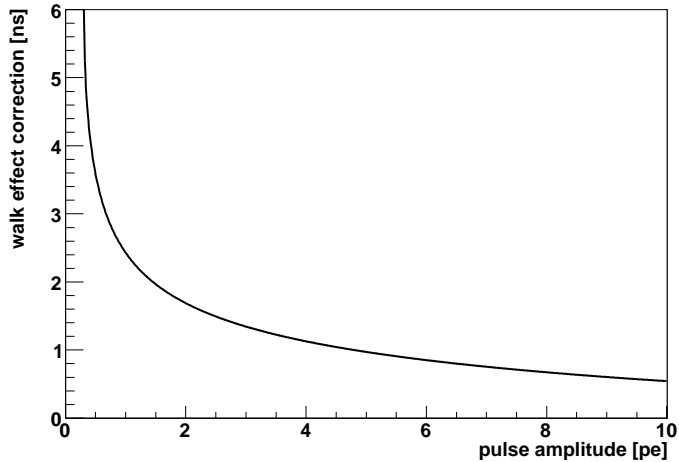}
\caption{\small Time offset due to the walk effect as a function of
the PMT pulse amplitude in units of photo-electrons.}
\label{fig:walk}
 \end{center}
\end{figure}

The fact that the PMT signals are discriminated by the ARSs using a
fixed amplitude threshold leads to a \textit{walk effect}: a PMT
signal with high amplitude crosses the threshold earlier than a
coincident low amplitude signal. The difference between the threshold
crossing times of pulses of different amplitudes can be of the order
of a few nanoseconds. This effect can be corrected if the
photo-electron pulse shape (PPS) is known: the PPS function is scaled
up or down to reproduce the total charge of the hit and the
corresponding threshold-crossing time is calculated.  During standard
data acquisition the total charge of a PMT hit is integrated by the
ARS and the information about its shape is lost. However, the ARSs can
be operated in a special acquisition mode which records PMT signals as
waveforms sampled into 128 channels over 200 ns and digitized by an
ADC. The PPS used to extract the walk correction is computed by
fitting the average waveform obtained from single photo-electron
events of all the ARSs.  The walk effect correction as a function of
the pulse amplitude is shown in Figure~\ref{fig:walk}. It is
systematically applied before event reconstruction.

\section{The Clock System}
\label{sec:clock}

The main goal of the clock system is to provide a common signal to
synchronize the readout of the OMs. It consists of a 20 MHz generator
on shore, a distribution network and a signal transceiver in each
LCM. The optical clock signals received from shore are converted and
decoded by an electronics board inside the LCM, and are then
distributed over an electrical bus to the ARS chips.

The absolute time stamping is performed by interfacing the clock
system with a dedicated electronics board to the GPS timing system
that provides a time accuracy of $\sim$100~ns with respect to the
Universal Time Coordinated, well within the requirements (a few
seconds to provide the direction of the neutrino in celestial
coordinates and a few milliseconds to correlate it to astrophysical
phenomena).

Track reconstruction requires the knowledge of the
relative arrival times of the Cherenkov photons at the PMTs and
therefore only their relative time offsets.

\begin{figure}
 \begin{center}
 \epsfxsize=14.0cm
  \epsffile{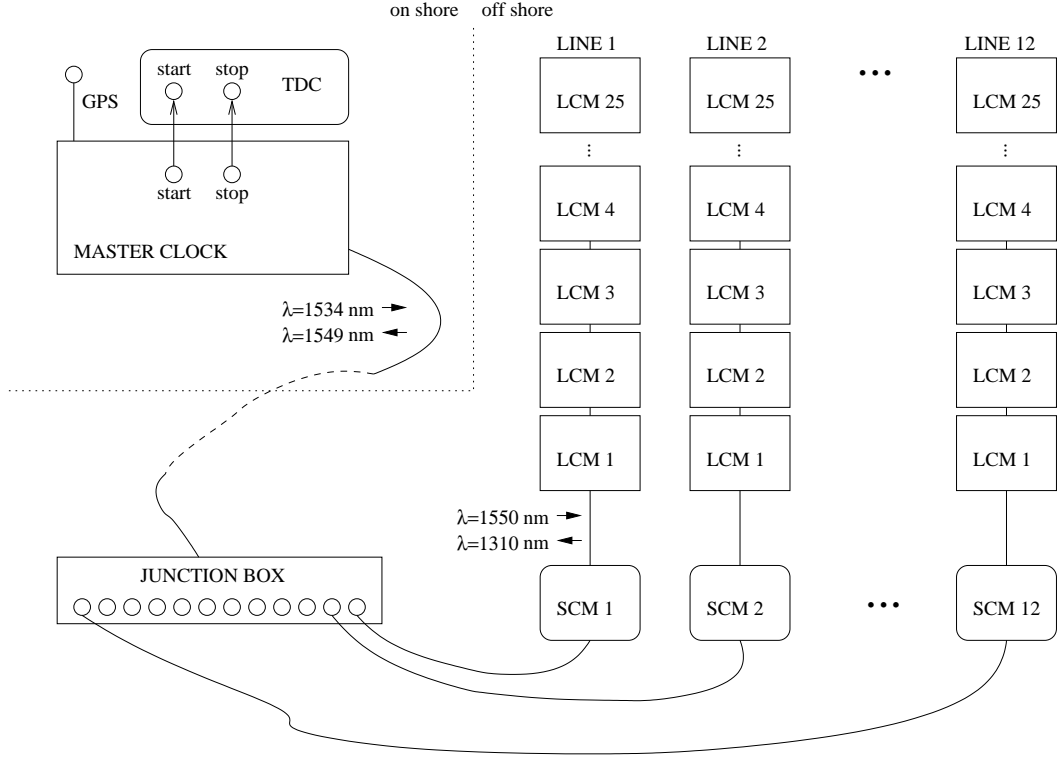} 
 \caption{\small Schematic view of the clock system: optical signals
 are sent from shore and propagated up to the junction box where they
 are split towards the SCMs of the different lines.  From the SCM they
 are distributed to each LCM, where a transceiver sends the signal
 back to shore for calibration purposes.}
 \label{fig:clock}
 \end{center}
\end{figure}

The layout of the clock distribution network is shown in
Figure~\ref{fig:clock}. It is based on a bi-directional optical
communication system. The clock signals are transmitted (received)
between shore and the SCMs of the lines using infrared signals of
1534~nm (1549~nm) wavelength, while within each detector line
wavelengths of 1550 nm and 1310 nm are used for transmissions in the
two opposite directions. In order to measure the time delays for
individual storeys due to the propagation time along the cables,
special clock signals are addressed to each LCM clock transceiver
which responds by sending back a signal to the shore. The
corresponding round-trip time is twice the propagation time to each
individual LCM. During the standard operation of the telescope, the
round-trip times to all the LCMs of the detector are monitored once
per hour, which ensures to have at least one measurement per data
taking run. An average of many such measurements provides the required
precision.

\begin{figure}
 \begin{center}
 \begin{tabular}{c c}
 \epsfysize=4cm
 \epsfxsize=9cm
 \epsffile{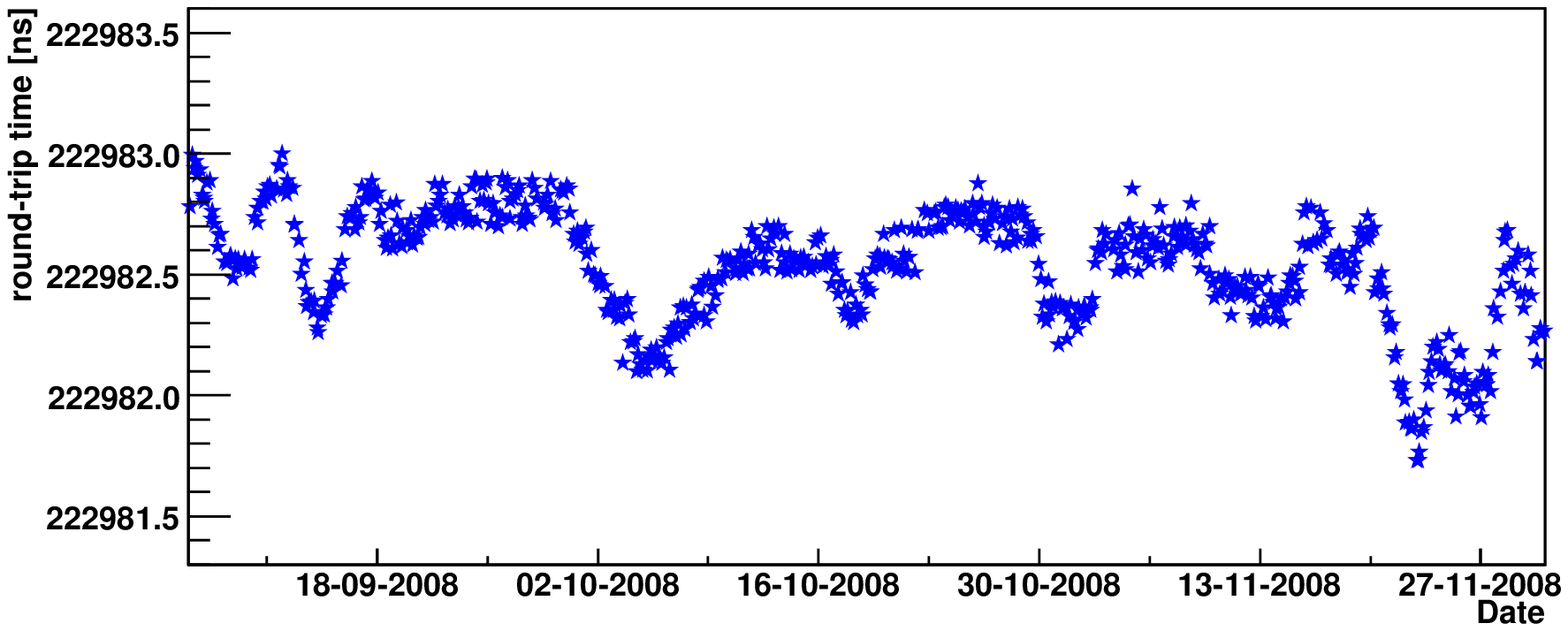}&
 \epsfysize=4cm
 \epsfxsize=5cm
 \epsffile{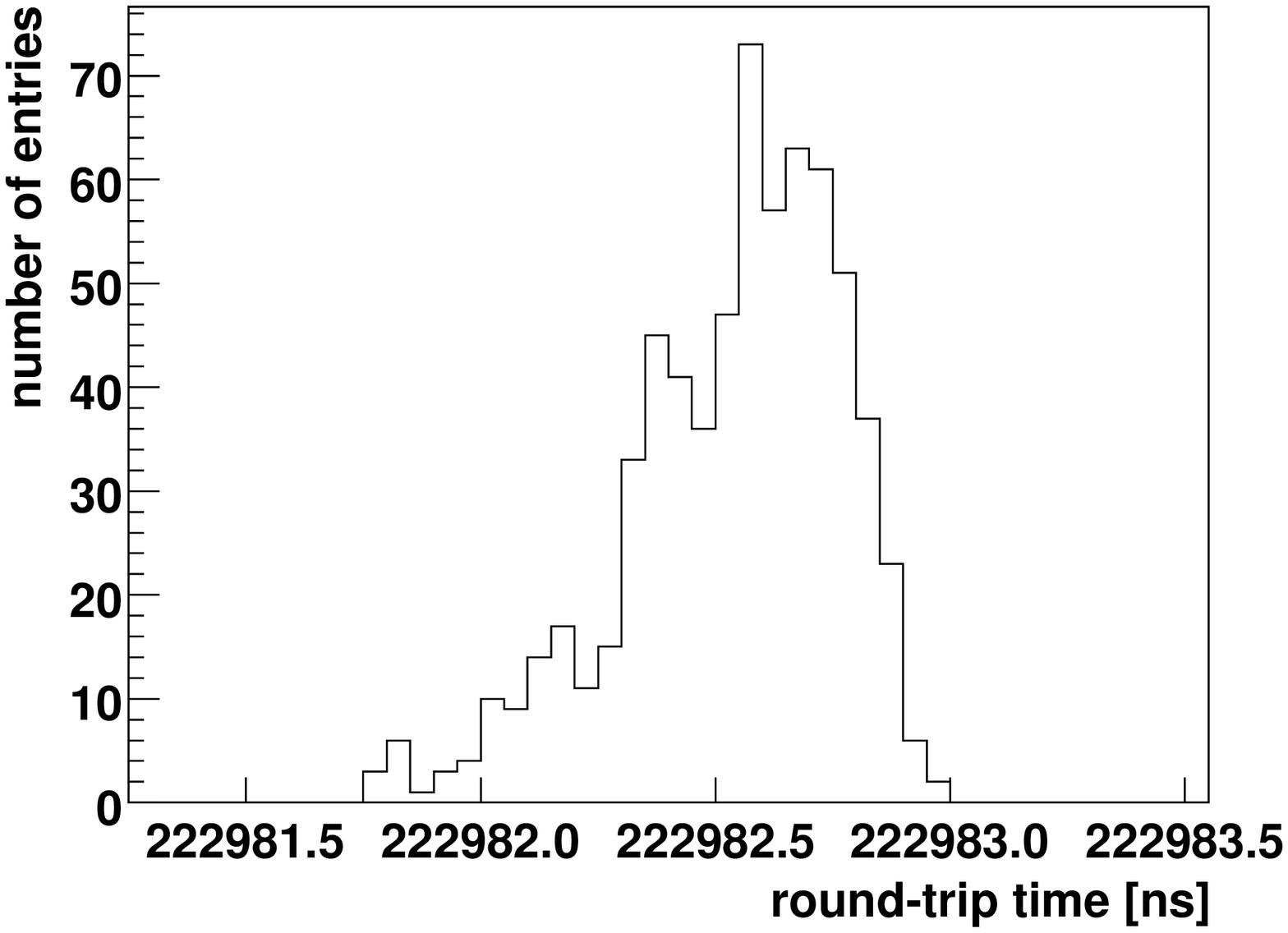} 
 \end{tabular}
 \caption{\small Left: Time evolution of the round-trip time of a
   clock signal from the onshore station to an SCM for the period
   August\--December 2008.  Right: The corresponding distribution of
   round-trip times for the same period.}
 \label{fig:evolution}
 \end{center}
\end{figure}

Figure~\ref{fig:evolution} shows an example of the time evolution of
the measured round-trip time from the shore station to the SCM of a
line. Clock delay variations of the order of a few hundreds of
picoseconds are observed and found to be correlated with temperature
variations in the onshore section of the MEOC.

\begin{figure}
 \begin{center}
 \begin{tabular}{c c}
 \epsfysize=4cm
 \epsfxsize=9cm
 \epsffile{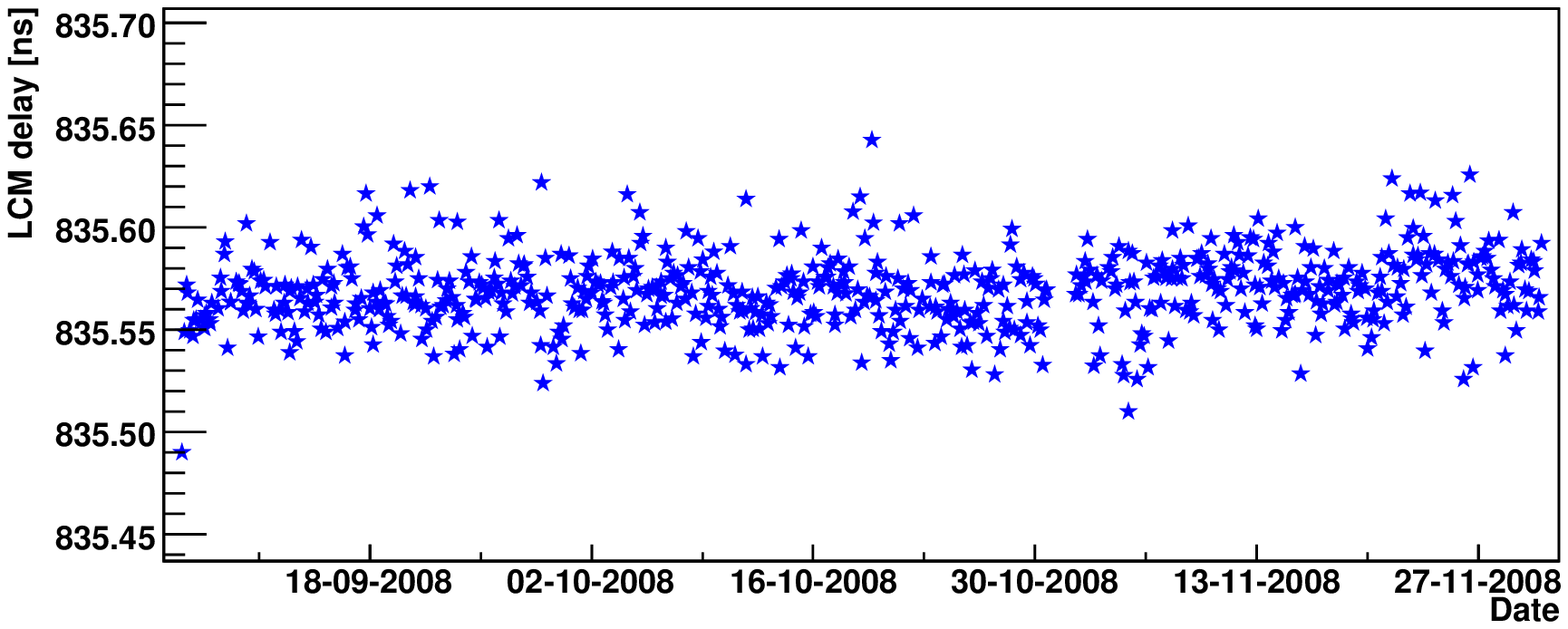}&
 \epsfysize=4cm
 \epsfxsize=5cm
 \epsffile{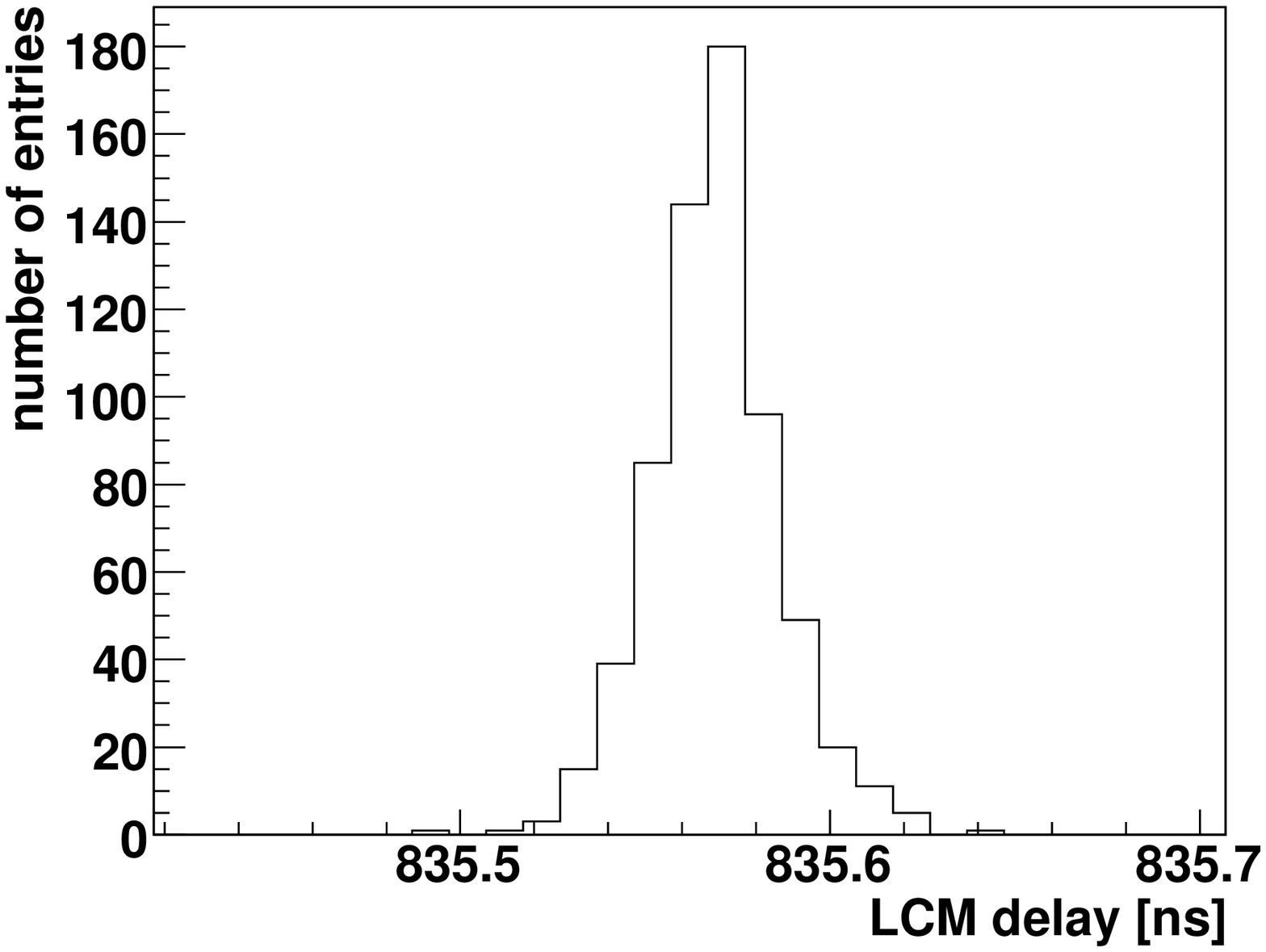}
 \end{tabular}
 \caption{\small Left: Time evolution of the difference in the
 round-trip times between the shore and an SCM and one of the
 LCMs. Right: Projection of these values, whose distribution has an
 RMS of about 16 ps.}
 \label{fig:roundtrips}
 \end{center}
\end{figure}

Figure~\ref{fig:roundtrips} shows the stability of the
round-trip times between an LCM and the SCM of a line. In this case,
the delay variations in the common part of the MEOC cable cancel out
and the relative uncertainty is reduced to around 16~ps.

\section{OM time offset calibration}
\label{sec:omtimes}
The time elapsed between the incidence of a photon on the photocathode
of the PMT and the time stamping of the associated signal in the ARS
must be determined. This delay depends on the PMT signal transit time,
the length of the cable linking the OM to the LCM, the front-end
electronics response time and the ARS threshold. The relevant time
constants are initially determined with a dedicated calibration set-up
in a {\it dark room} prior to deployment. These constants could change
slightly after deployment due to the temperature changes, stressing of
cables during transportation/deployment, etc. Morever, when the
high voltage applied to the PMTs change, the time offsets change
too. Using the optical beacon system, they are periodically monitored
and recomputed when the lines are operated in the sea.

\subsection{Onshore calibration}
\label{sec:darkroom}
Before deployment the functionality of each line is verified and a
complete calibration is performed. The onshore time calibration is
carried out illuminating simultaneously groups of OMs by short laser
pulses. The propagation times from SCM to LCM are known from the clock
system calibration. The contribution from the cable that links the LCM
to the OM, the PMT transit time and the front-end electronics delay
can thus be extracted.

\begin{figure}
 \begin{center}
 \epsfxsize=10.0cm
\epsffile{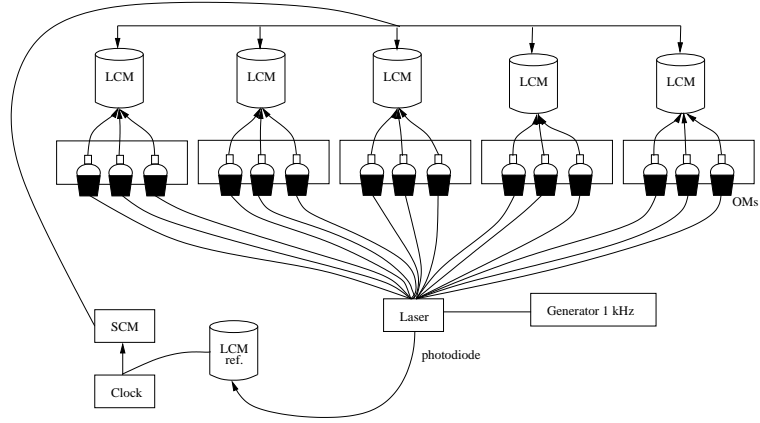}
 \caption{\small Set-up for time calibration in the dark room. A laser
 signal is split and sent to the 15 OMs of the sector to be calibrated.}
 \label{fig:drsetup}
 \end{center}
\end{figure}

The time calibration is performed in batches of five storeys (a
sector) in the dark room (see Figure~\ref{fig:drsetup}).  The light
source used is a Q-switched, Nd-YAG laser~\cite{bib:nanolase} that
emits intense (E$\sim$1~$\mu$J) and short (FWHM$\sim$0.8~ns) pulses of
green light ($\lambda$=532~nm) externally triggered at a frequency of
1~kHz.  The light is split using a 1-to-16 optical splitter and
distributed to the 15 OMs of each sector through optical fibres.  Each
optical fibre is coupled to a Lambertian diffuser which spreads the
laser light over the full area of the corresponding PMT photocathode.
The output amplitude of the light (equivalent to about 10$^{12}$
photons) is attenuated in order to produce a few tens of
photoelectrons in the PMTs. The fibre lengths are adjusted such that
the corresponding time differences are less than 0.3~ns. Measurements
with the optical fibres swapped between OMs are made to cross-check
these time differences.  The time reference is provided by the laser
internal photodiode. A reference LCM digitises this photodiode signal
in order to synchronize the time measurements for OMs of different
sectors.

\begin{figure}
 \begin{center}
 \epsfxsize=10.0cm
\epsffile{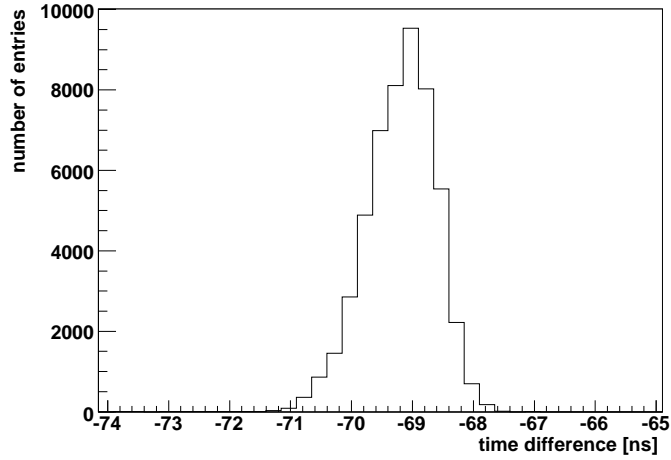}
 \caption{\small An example of the distribution of the time difference
   between the signal in one OM and the signal given by the reference
   photodiode of the laser during the dark room calibration.}
 \label{fig:timediff}
 \end{center}
\end{figure}

 An example of the distribution of the difference between the time of
 the signal recorded by an OM and the emission time of the pulse as
 recorded by the laser photodiode is shown in
 Figure~\ref{fig:timediff}. The mean value of this distribution yields
 the difference between the laser light signal time measured by the
 PMT (including the PMT transit time and the electronics delay) and
 the time measured by the photodiode channel. The sigma of the
 distribution is 0.5~ns.

\begin{figure}[!ht]
 \begin{center}
 \epsfxsize=10.0cm
 \epsffile{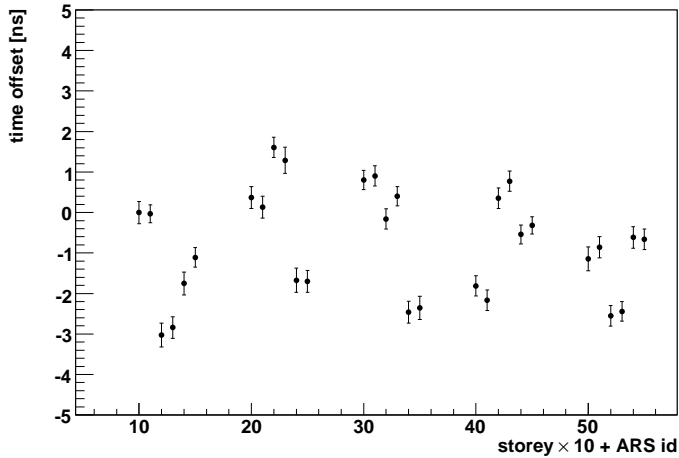}
 \caption{\small Time offsets measured for all ARSs of one of the line
 sectors. The time offsets are relative to the first ARS of the first
 OM of the lowest storey. The ARS id runs from 0 to 5 within each
 storey.}
 \label{fig:line1}
 \end{center}
\end{figure}

The time offsets between each OM of a line and the first OM of its
lowest storey, which is chosen as a reference, are computed for each
of the 30 ARSs of each sector. An example result is shown
in Figure~\ref{fig:line1}. The spread of the time offsets is of the order of
a few nanoseconds, and is due to differences in the internal cabling
of the OMs to the storey electronics as well as differences in PMT
transit times. These time offsets are stored in the ANTARES database
and are used as initial time constants in order to analyse data
immediately after the installation of the line on the sea bed.

The relative time calibration of the monitoring PMTs of the LED
optical beacons (used for the {\it in situ} calibration described in
subsection~\ref{sec:lobs}) is also performed during the onshore
calibration.  One of the fibres of the optical splitter is used in
this case to send the light to the small PMT within each OB. Since
this PMT is meant to measure the relatively high intensity of light
emitted by the LEDs, the incident laser light intensity is tuned to be
about twenty times higher than that provided to the OMs.

\subsection{{\it In situ} calibration}
\label{sec:insitu}

\begin{figure}
 \begin{center}
 \begin{tabular}{c c}
 \epsfysize=4.8cm
 \epsffile{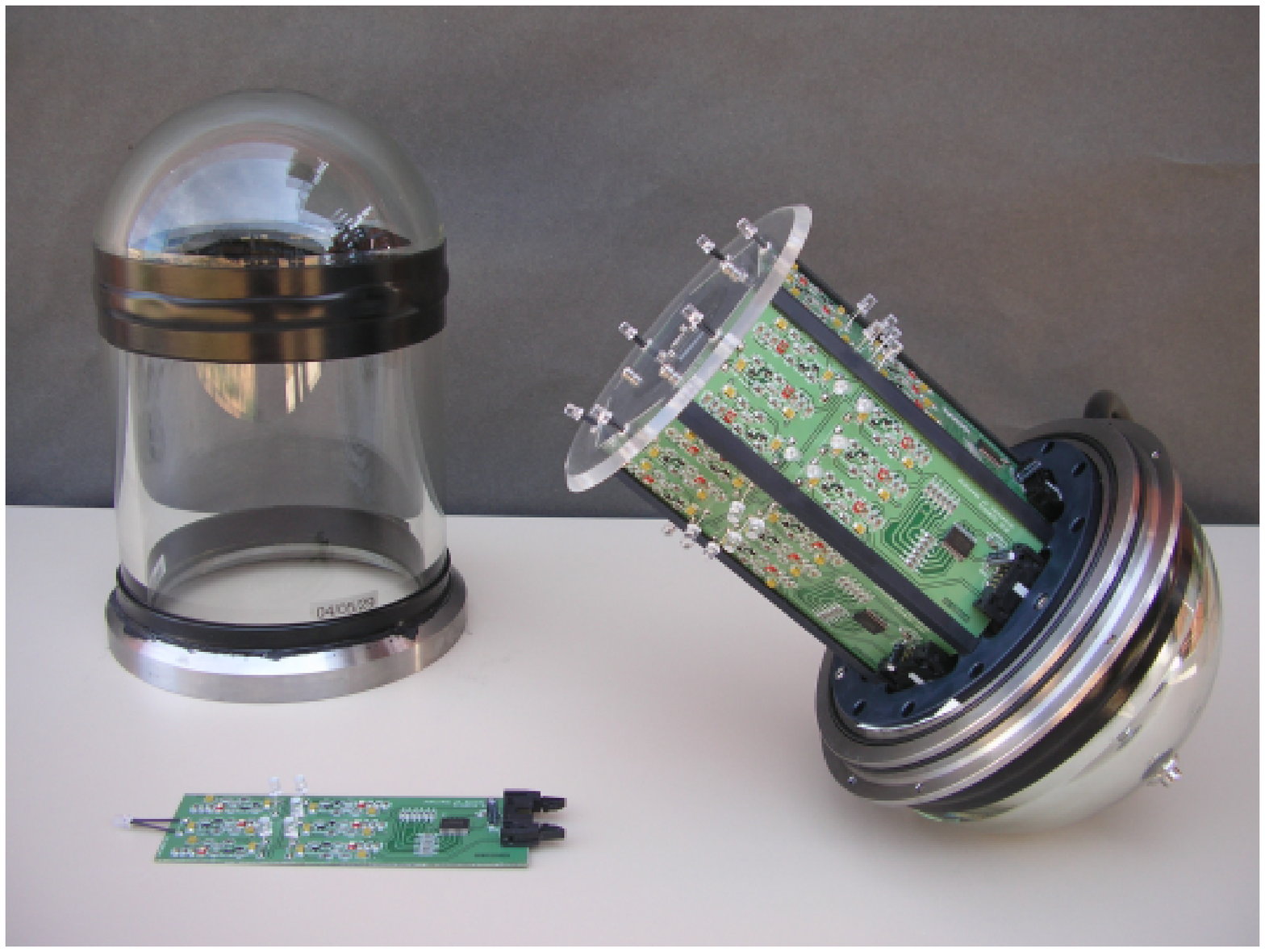}&
 \epsfysize=4.8cm 
\epsffile{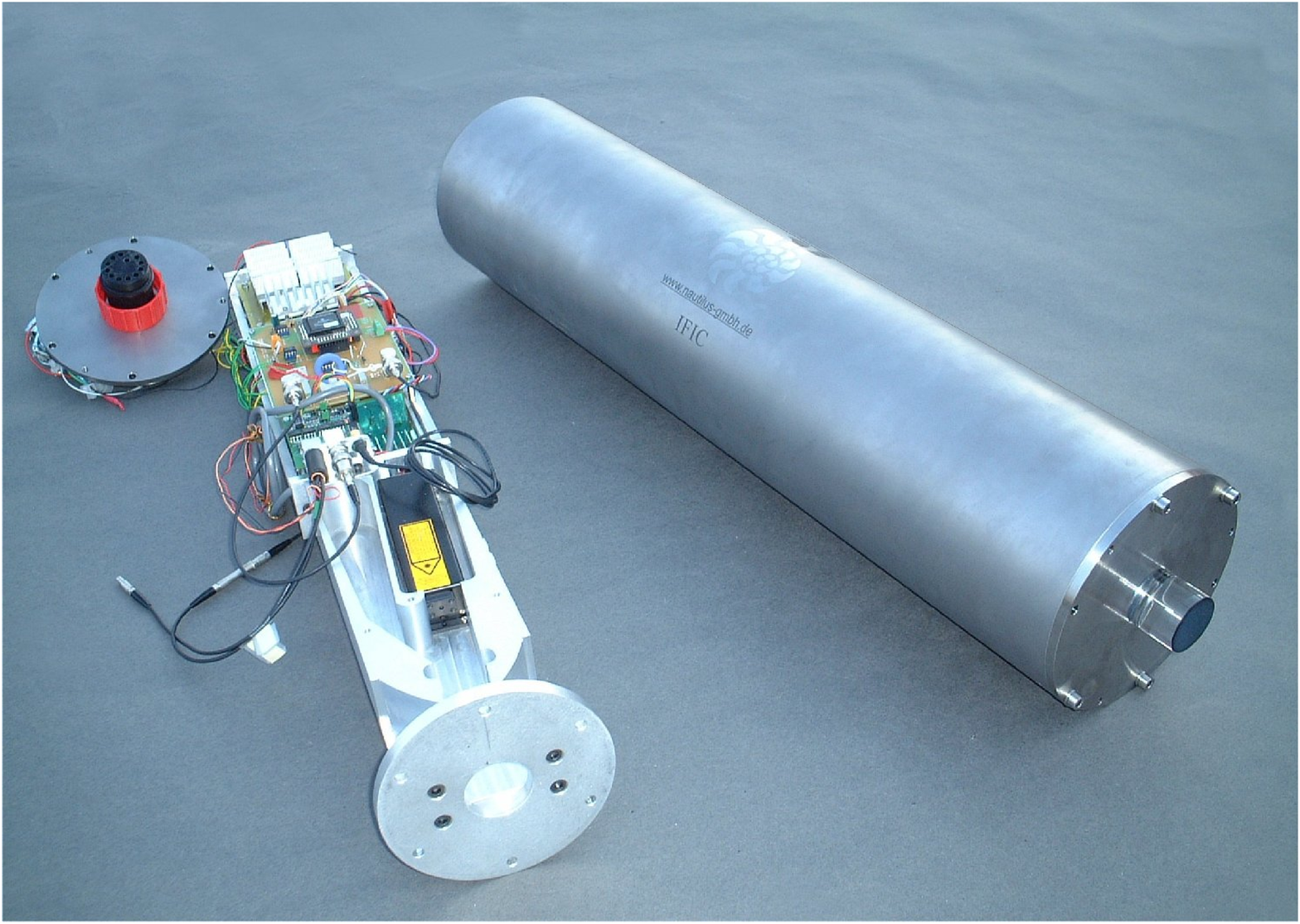} 
 \end{tabular}
 \caption{\small Picture of an LED optical beacon (left) and a laser
 beacon (right).}
 \label{fig:obs}
 \end{center}
\end{figure}

The {\it in situ} measurement of the time offsets of all the OMs is
performed with the optical beacon system~\cite{bib:obpaper}.  This
comprises two kinds of complementary devices: LED beacons that emit
blue light ($\lambda=470$~nm) and laser beacons emitting green light
($\lambda=532$~nm) (see~\cite{bib:ompaper} for a description of the
response of the OM as a function of the wavelength). Pictures of both
devices before assembly are shown in Figure~\ref{fig:obs}. The LED
beacons are used to monitor the relative time offsets among OMs of the
same line (intra-line calibration), whereas the laser beacons are used
for monitoring the relative time offsets among the lines (inter-line
calibration) and the calibration of the lowest storeys. The {\it in
situ} calibration allows to re-measure the time constants computed
onshore and is particularly important in case of a change in the PMT
high-voltage and threshold settings.

\subsubsection{LED beacons}
\label{sec:lobs}

Four LED optical beacons are located along every line of the detector
in storeys 2, 9, 15 and 21 (counted from bottom to top). Each LED OB
contains 36 LEDs distributed on six faces forming a hexagonal
prism. The LEDs emit blue light with a maximum intensity of
$\sim$160~pJ per flash ($\sim$4$\times 10^8$ photons per pulse) and a
pulse width of 4~ns (FWHM). The emission time of the LED light is
measured by an 8-mm PMT placed inside the frame of
the LED OB (see reference~\cite{bib:obpaper} for more details).

\begin{figure}
 \begin{center}
 \begin{tabular}{c c}
 \epsfysize=7.0cm
 \epsffile{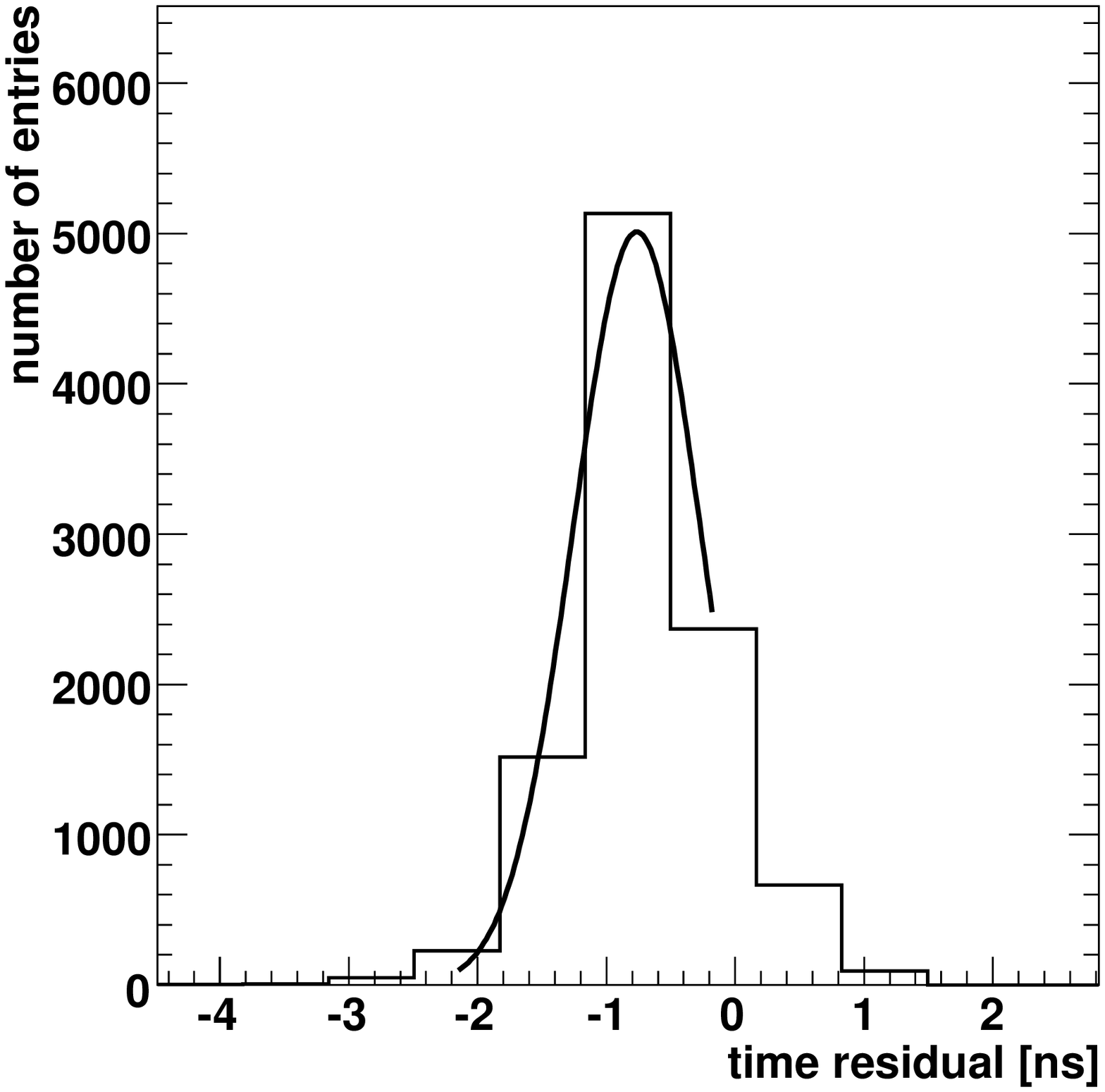}&
 \epsfysize=7.0cm
 \epsffile{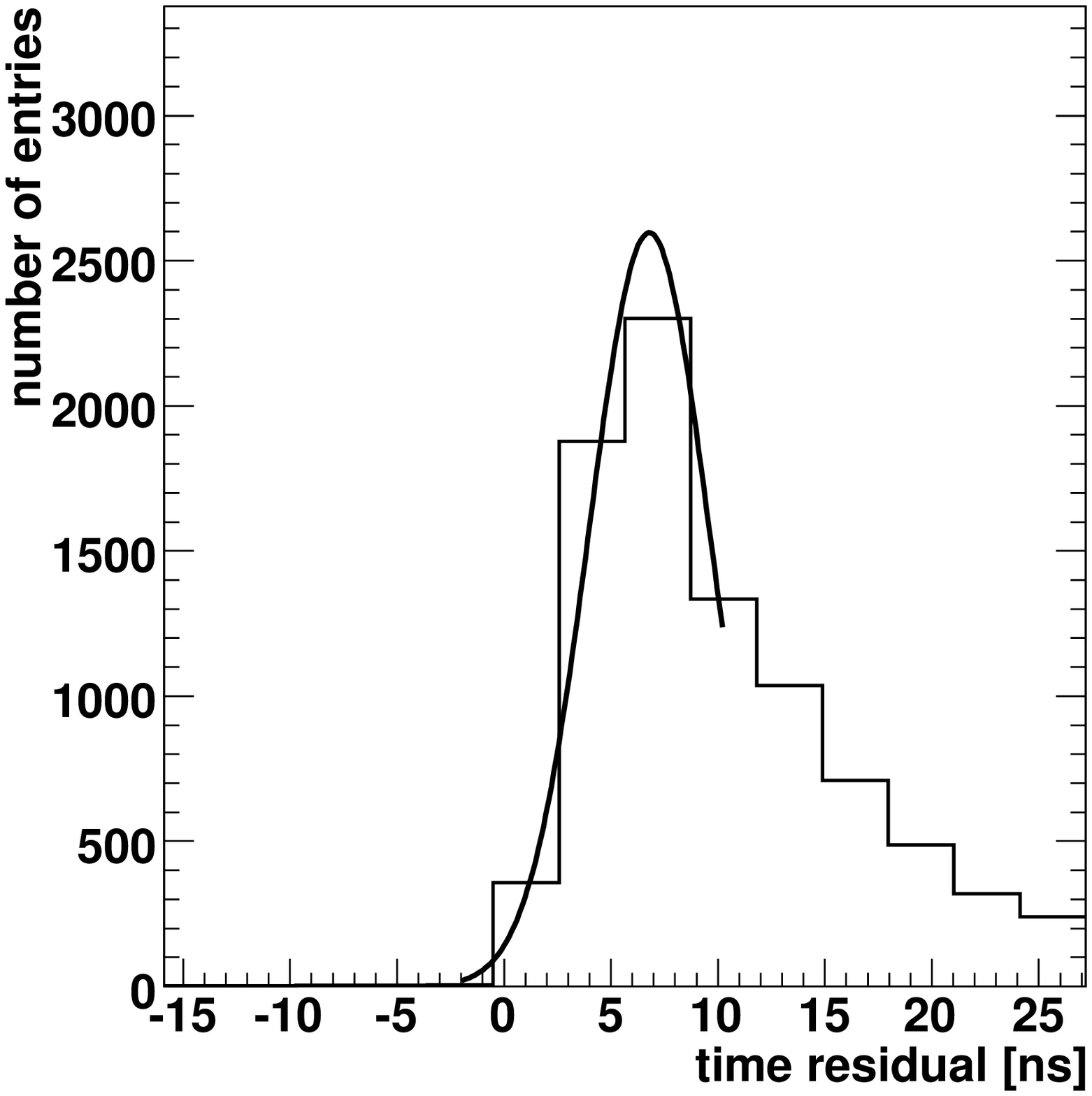}
 \end{tabular}
 \caption{\small Time residual distribution in two different OMs
 located two storeys (left) and seven storeys (right) above the LED
 flashing beacon. The curve is the Gaussian fit explained in the
 text.}
 \label{fig:timedis}
 \end{center}
\end{figure}

Figure~\ref{fig:timedis} shows the distribution of signal time
residuals, i.e. the difference between the signal time recorded in the
OM and the emission time of the corresponding LED flash, using the
time offsets measured in the dark room and once the nominal travel
time of the light from the beacon to the OM has been subtracted. The
plot on the left shows the distribution corresponding to an OM close
to the OB ($\sim$30~m) which receives a high amount of light. The plot
on the right shows the distribution for an OM located seven storeys
above the OB ($\sim$100~m). In this case, a tail at positive times due
to light scattering in water can be observed. A Gaussian fit to the
region including the rising edge of the time distribution and the
first bin after its maximum is performed. The range of the fit is
restricted in order to use mostly the earliest photons, whose delay
due to scattering can be neglected.

\begin{figure}
 \begin{center}
 \begin{tabular}{c c}
 \epsfysize=4.8cm \epsffile{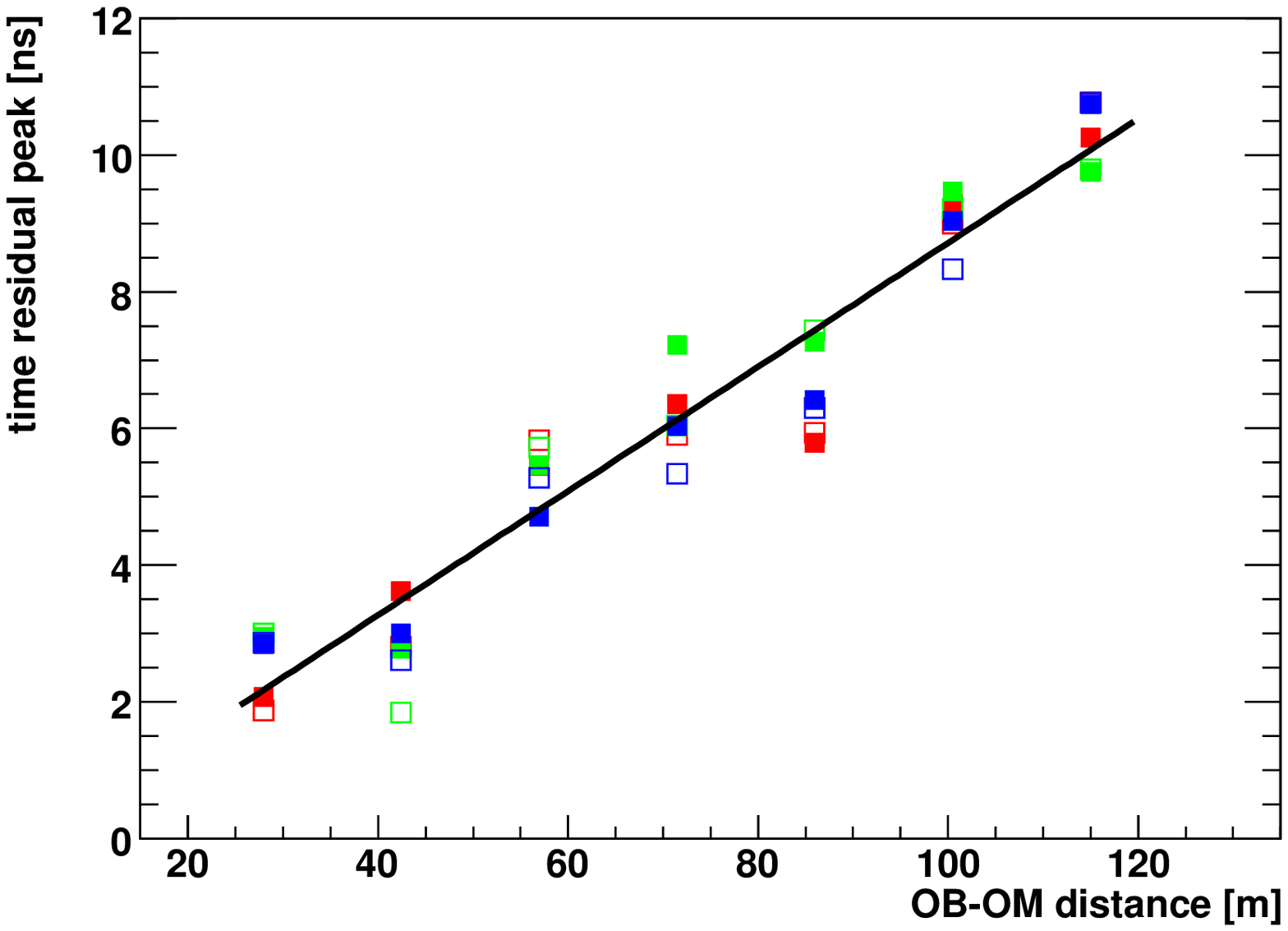}&
 \epsfysize=4.8cm \epsffile{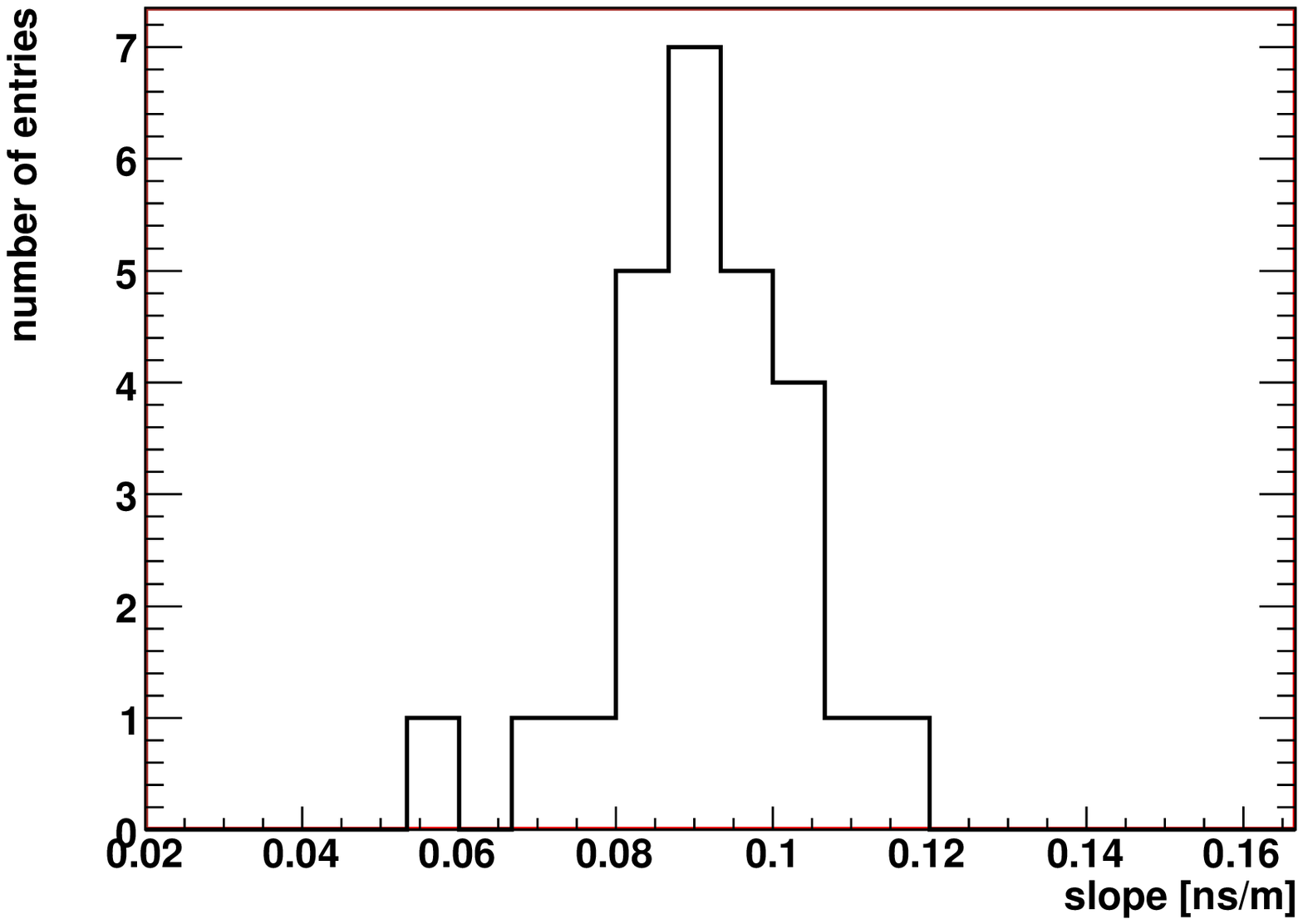}
 \end{tabular}
 \caption{\small Left: Time residual peak position as a function of
 the OB-OM distance, i.e. the distance between the optical beacon and
 the different optical modules, for all the ARSs along seven
 storeys. The six points at each distance correspond to the six ARSs
 in each storey (two ARSs per OM). The additional delay with distance
 is due to the early photon effect (see text). Right: Distribution of
 the slopes of the fits to the time residual peak positions as a
 function of the OB-OM distance for all the OBs.}
 \label{fig:recta}
 \end{center}
\end{figure}

\begin{figure}
 \begin{center}
 \epsfysize=6.0cm
 \epsffile{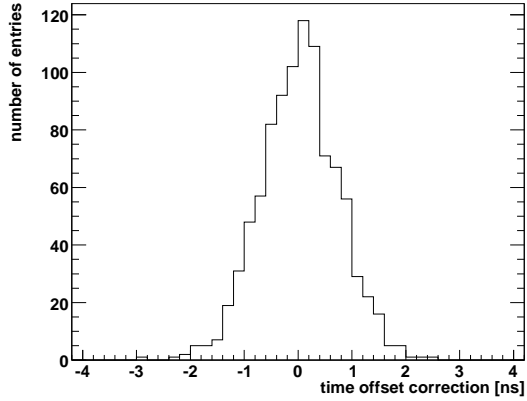}
 \caption{\small Distribution of the time offset corrections measured
 {\it in situ} with the LED optical beacons. Each entry is calculated
 as the difference between the time residual peak shown in
 Figure~\ref{fig:recta} and the corresponding value of the fitted
 line.}
 \label{fig:corrLOB}
 \end{center}
\end{figure}

The position of the time residual peak, i.e. the mean value of the
Gaussian best-fit curve described above, as a function of the distance
from the OM to one optical beacon is shown in Figure~\ref{fig:recta},
together with the slopes of such fits for all the OBs. The increase of
the time residual with distance originates from an ``early photon''
effect, arising from the duration of the light pulse and the fact that
the first photons detected by the PMT determine the recorded time of
the pulse~\cite{bib:juanan}.  The PMTs closer to the beacon, which
detect a higher amount of photons per pulse, tend to trigger on the
photons emitted earlier in the flash, whilst for PMTs farther away the
probability distribution for the measured time of the signal is
determined by the pulse width. The arrival time of the first photon
obeys the {\it order statistics}~\cite{bib:kendall}. A Monte Carlo
simulation taking into account this effect as well as the exponential
decrease of the intensity caused by the light attenuation shows that a
linear increase in the time residual is expected when the number of
photons reaching the PMTs is sufficiently high. A straight line is
fitted to the time residual peaks ordered by the distance of the
optical modules from the optical beacon. If a point deviates more than
2~ns, the fit is redone excluding it. The deviations from the fit are
then used to obtain the time offset correction. The distribution of
these time offset corrections for all OMs that can be calibrated with
the OBs is shown in Figure~\ref{fig:corrLOB}. It is found that 15\% of
the cases need a correction to the dark room offsets larger than 1~ns.

Seven storeys above each OB, excluding the one just above the beacon
(which receives too much light), can be calibrated. The storeys farther
away, which do not receive enough light, are monitored with the next
OB along the line. The OMs of the first three storeys of each line can
not be calibrated with the LED OBs since they are below or just above
the lowest OB. For these, the time offsets given by the dark room
measurements are used, corrected by the laser beacon and potassium-40
measurements as explained later.

\begin{figure}
 \begin{center}
 \begin{tabular}{c c}
 \epsfxsize=7cm
 \epsffile{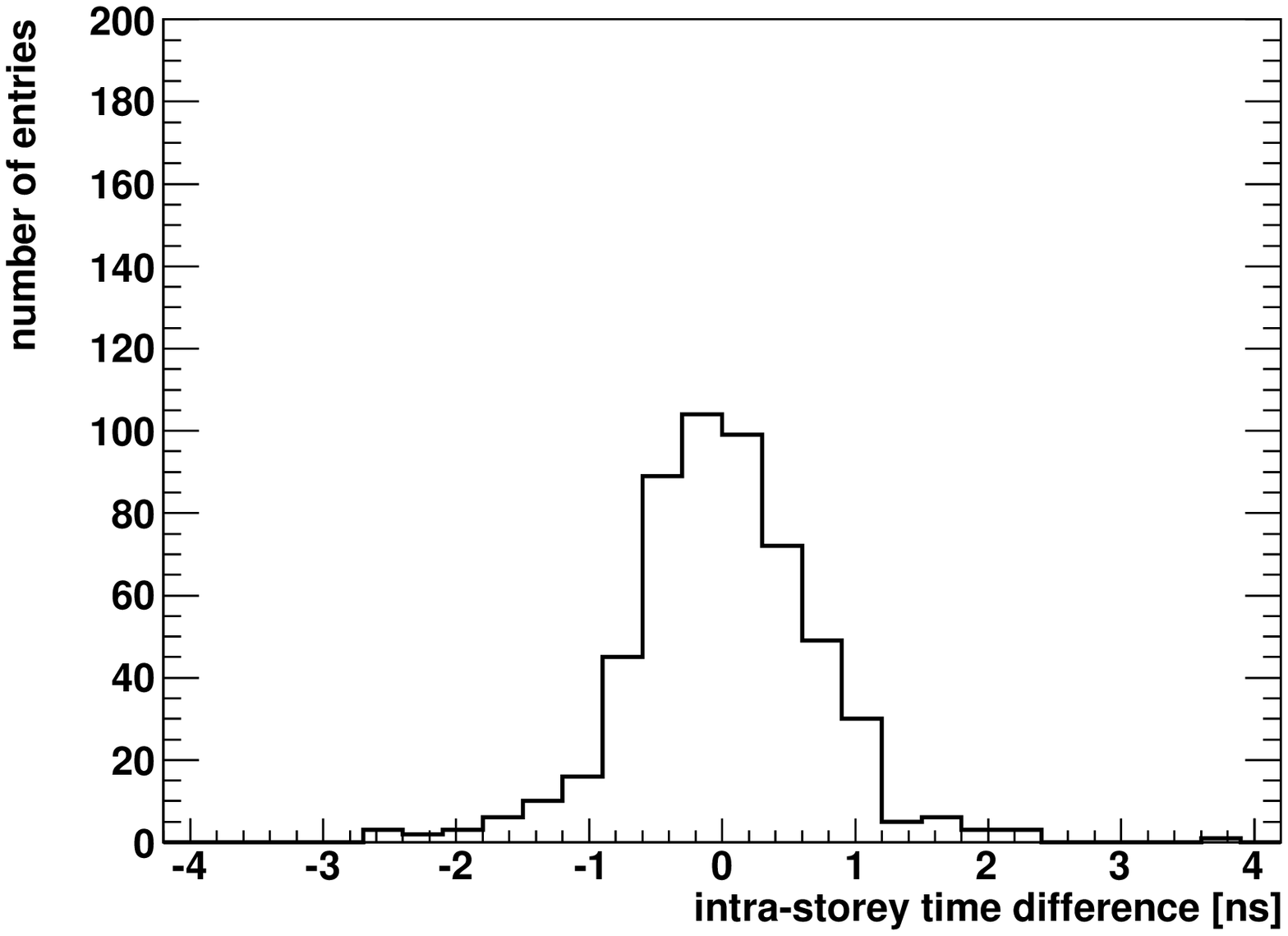} &
 \epsfxsize=7cm 
 \epsffile{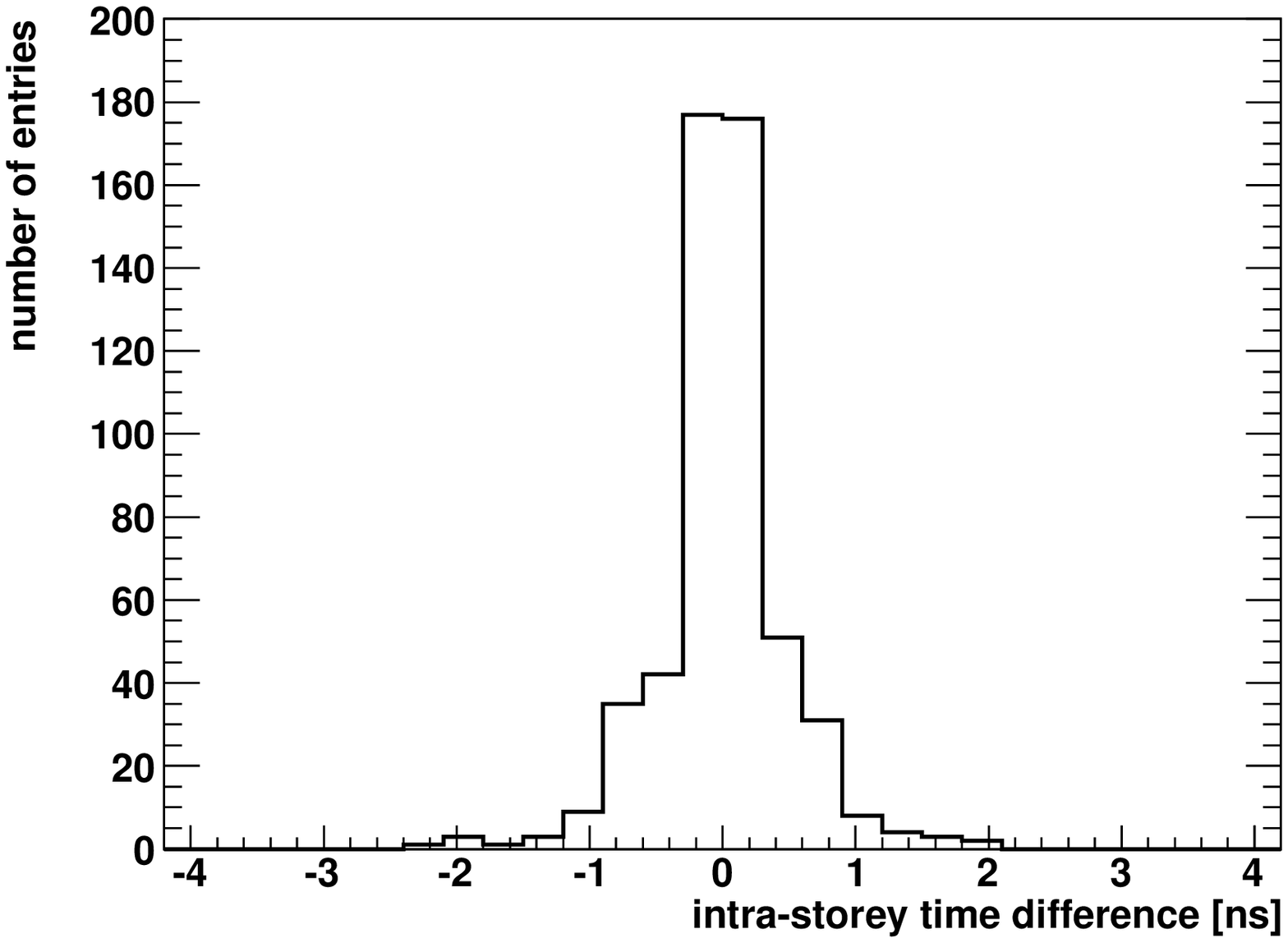}
\end{tabular}
 \caption{\small Time differences between OMs in the same storey for
 one of the lines using the time offsets measured onshore (left) and {\it
 in situ} (right).}
 \label{fig:intrastorey}
 \end{center}
\end{figure}

A way to check the validity of a set of time offsets is to calculate
the signal time differences for pairs of OMs in the same storey
(intra-storey time differences) when illuminated by the optical
beacon. In Figure~\ref{fig:intrastorey}, the intra-storey time
differences for one of the lines using the time offsets measured
onshore and {\it in situ} are shown. The time differences become
smaller when using the {\it in situ} offsets. For a detector
configuration when 10 lines were installed, the RMS of the average
intra-storey time difference distribution for the OMs which can be
re-calibrated decreases from 0.72~ns to 0.48~ns after the application
of the {\it in situ} calibration. This is a 30\% improvement, which is
equivalent to a subtraction of 0.54~ns in quadrature. In terms of
sigma, the change is from 0.60~ns to 0.33~ns. Initially calibration
runs were performed once per week, but after the time offsets were
found to be very stable, the frequency has been reduced to once per
month. Since each set of calibration runs takes about two hours, the
total dead time is about one day per year, which is considered short
enough.

\subsubsection{Laser beacon}
\label{sec:laser}

\begin{figure}
 \begin{center}
 \epsfxsize=10cm 
 \epsffile{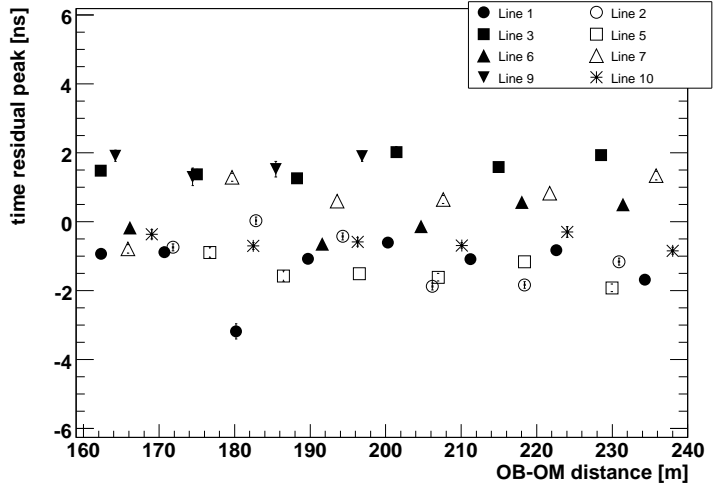}
 \caption{\small Time residual peak position versus the distance
 between the laser beacon and the OM. Each point is the average over
 all the OMs in the same storey.}
 \label{fig:interline}
 \end{center}
\end{figure}

\begin{figure}
 \begin{center}
 \epsfxsize=10cm 
 \epsffile{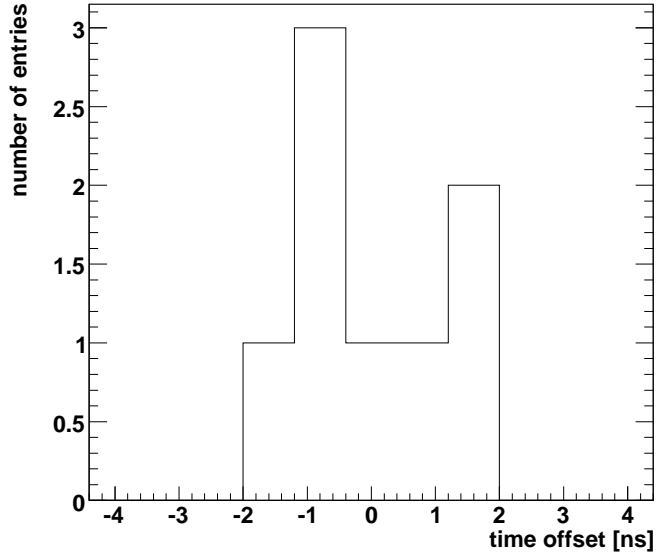}
 \caption{\small Distribution of the inter-line offsets calculated
 from the measurements made with the laser beacon. Each offset is
 calculated as the average of the time residual peaks shown in
 Figure~\ref{fig:interline}.}
 \label{fig:corrInterLine}
 \end{center}
\end{figure}

The relative time offsets among lines can be obtained with the laser
beacons. This inter-line calibration is needed because the time
offsets measured in the dark room are calculated with respect to a
reference OM specific to each line. Being much more powerful than the
LED beacons, the lasers can illuminate all detector lines. Given that
only distances where the OMs are illuminated below the
one-photoelectron level are considered and since the time width of the
laser pulse is very narrow (FWHM$\sim$0.8~ns), the early photon effect
is negligible. As a consequence, the time residuals do not depend on
the distance to the source, as can be seen in
Figure~\ref{fig:interline}.  The relative time offsets among lines are
then computed as the average of the time residual peaks. The
calculated inter-line time offsets are shown in
Figure~\ref{fig:corrInterLine}. In addition to the inter-line
calibration, the laser beacon also provides a tool to compute the time
offsets of the OMs in the lowest storeys.

\subsection{Potassium-40}
\label{sec:k40}

\begin{figure}
 \begin{center}
 \begin{tabular}{c c}
 \epsfysize=6cm 
 \epsffile{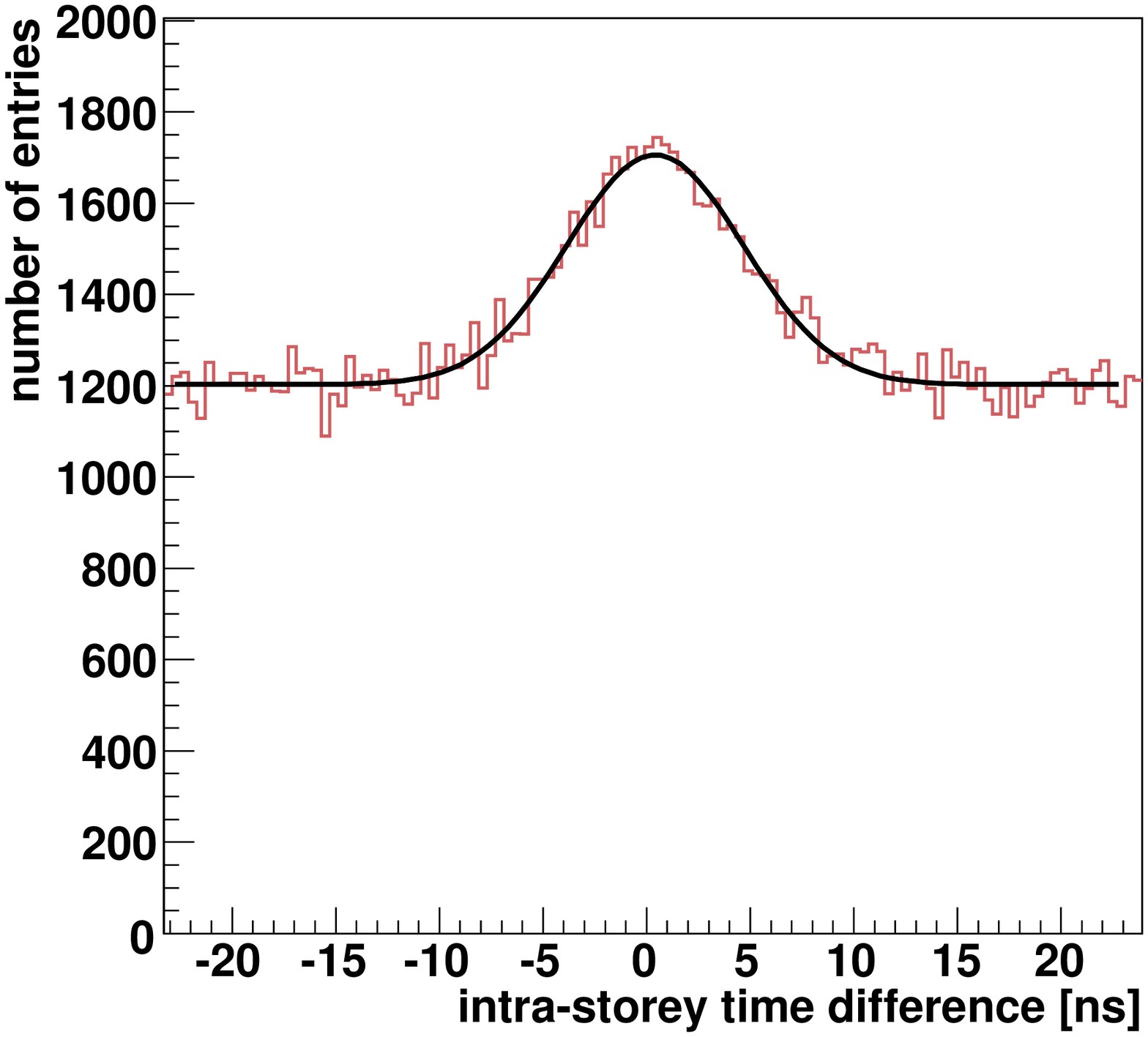} &
 \epsfysize=6cm 
 \epsffile{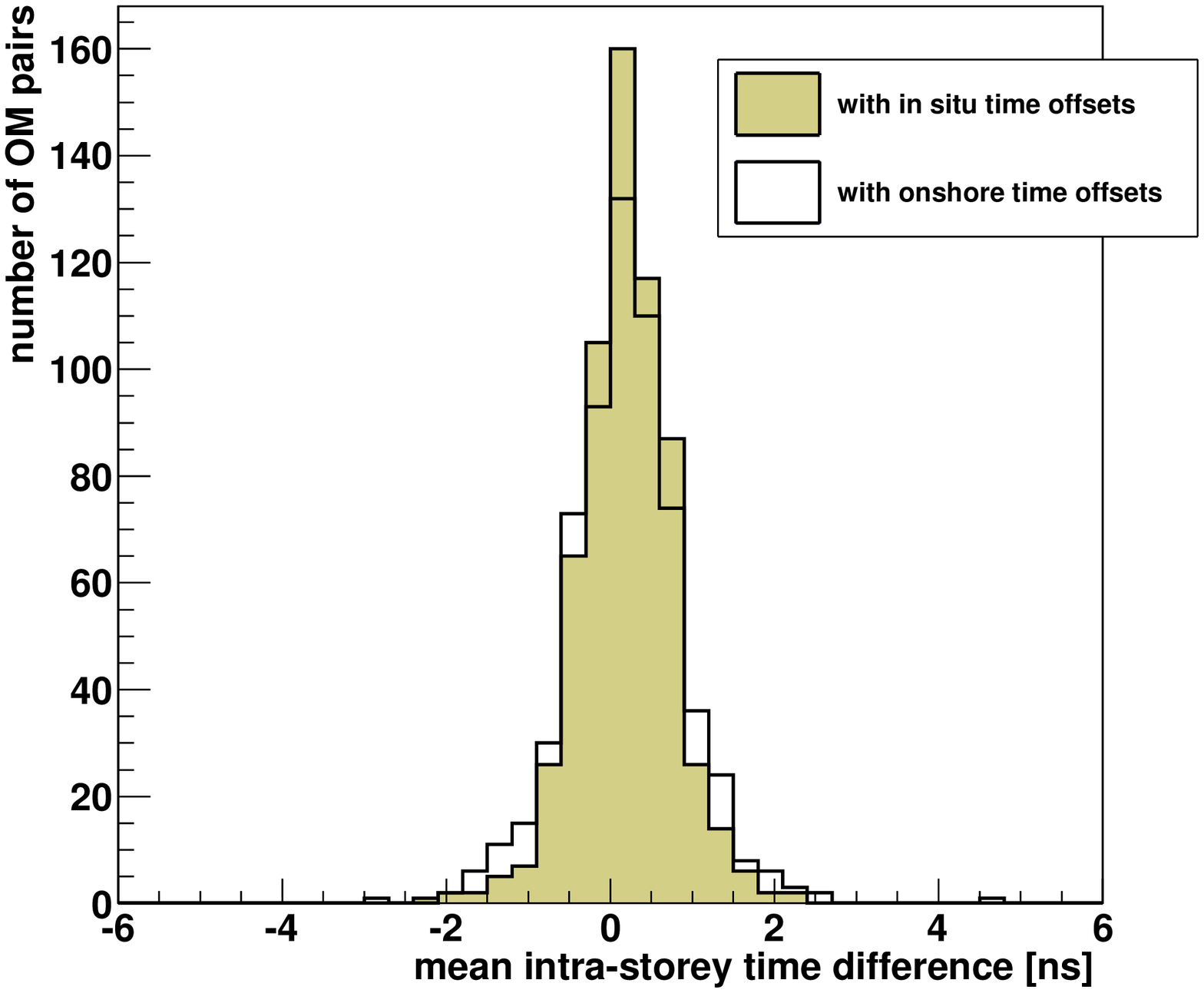}
 \end{tabular}
 \caption{\small Left: Distribution of background hit time differences
   for one pair of OMs in the same storey. The peak is due to single
   $^{40}$K decays detected in coincidence by two OMs. The data have
   been fitted to a sum of a Gaussian distribution and a flat
   background from random coincidences. Right: Comparison of the
   distributions of mean intra-storey time differences from $^{40}$K
   using the onshore and {\it in situ} time offsets.}
\label{fig:potassium}

 \end{center}
\end{figure}

The radioactive potassium-40 ($^{40}$K) present in sea water can be
used for time and efficiency calibration of the detector using the
Cherenkov light induced by the electron emitted in the $\beta$-decay
of potassium~\cite{bib:dmitry,bib:biolum}. If such a decay occurs
within a distance of a few meters from a storey, coincident signals
can be recorded by two OMs. In Figure~\ref{fig:potassium} (left) an
example of the distribution of the measured time differences between
hits in two OMs of the same storey is shown. A clear peak is found, in
good agreement with the expectations from simulations. The data has
been fitted to the sum of a Gaussian distribution and a flat
background. The width of these distributions (FWHM=9~ns) is set by the
difference in the distance from the point where the decay occurs to
each of the OMs of the pair. The position of the peak can be used to
cross-check within each storey the time offsets provided by the
onshore dark room and optical beacon calibrations
(Figure~\ref{fig:potassium}, right). If the time offset of one of the
OMs of the pair were incorrect, we would see that the peak is
displaced from zero. The RMS of the mean intra-storey time difference
distribution determined by the $^{40}$K improves from 0.72~ns to
0.57~ns when using the time offsets calculated {\it in situ} rather
than those determined from the dark room calibration. It is worth
noticing that the $^{40}$K intra-storey calibration is independent of
the LED OB system and relies on a completely different light source:
the $^{40}$K is a dim, distributed and closeby source, whereas the
beacons are powerful, point-like and distant sources.

\subsection{Internal LEDs}
\label{sec:internal}

\begin{figure}
 \begin{center}
 \epsfxsize=10.0cm
 \epsffile{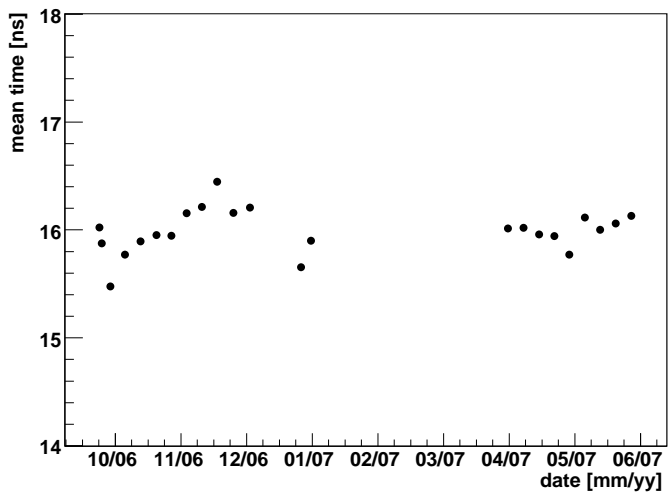}
 \caption{\small Example of the mean time of the internal LED flashes
 as recorded by the PMT as a function of the time.}
 \label{fig:internal}
 \end{center}
\end{figure}

Each optical module also incorporates an internal LED (of the same
model as those used in the LED beacons) which is used exclusively to
monitor the stability of the PMT transit time. The LED is located on
the back of the phototube and illuminates the photocathode from
behind. It is triggered by the clock signal at a constant rate.
Figure~\ref{fig:internal} shows the mean time of the internal LED
flashes as recorded by the corresponding PMT as a function of
time. The values, measured {\it in situ} every week, vary less than
0.2~ns (RMS) over an eight month period.

\section{Time resolution of the front-end electronics}
\label{sec:timeres}

The impact of the front-end electronics on the time resolution can be
determined during onshore and {\it in situ} calibrations. In this
section we present three different methods to estimate this front-end
electronics contribution.

 At low light intensity, the time resolution of an OM measured in the
laboratory is dominated by the transit time spread of the PMT
($\sigma_{TTS} \sim 1.3$~ns). At high intensity, this contribution
decreases as the square root of the number of photoelectrons and
therefore the dominant term to the width of this distribution is the
constant contribution due to the front-end electronics. This
irreducible contribution is found to be $\sim$0.5 ns in the dark room
calibration (see Figure~\ref{fig:timediff}).

\begin{figure}
 \begin{center}
 \epsfxsize=10.0cm
 \epsffile{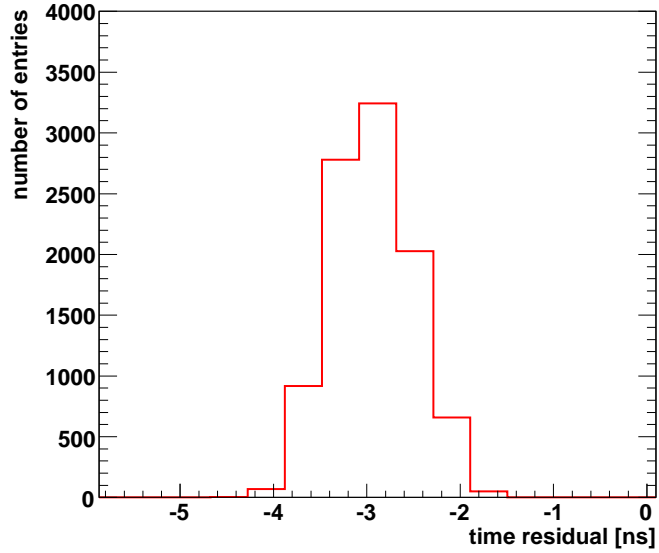}
 \caption{\small Example of the time residual distribution for an OM
located one storey above the optical beacon. The sigma of the
distribution (0.5~ns) is dominated by the contribution from the
front-end electronics.}
 \label{fig:closeom}
 \end{center}
\end{figure}

The {\it in situ} estimation of the time resolution of the electronics
is obtained from the distribution of the difference in the time
measured by an OM close to an OB with respect to the emission time of
the OB pulse. The sigma of the time distribution of the signal in an
OM is given by

\begin{equation}
  \sigma^{2}_{OM} = \frac{\sigma^{2}_{TTS}}{N_{pe}} + \frac{\sigma^{2}_{water}}{N_{\gamma}} + \sigma^{2}_{OB} + \sigma^{2}_{elec}
\label{eq:resol}
\end{equation}
\par\noindent where $\sigma_{TTS}$ is the transit time spread of the
PMT, $\sigma_{water}$ is the spread due to the scattering and
chromatic dispersion of light in water ($\sim$1.5~ns for a light path
of 40~m), $\sigma_{OB}$ is the uncertainty of the measured emission
time of the pulse and $\sigma_{elec}$ the spread due to the
electronics. For an OM close to an OB, $N_{pe}$ and $N_{\gamma}$ are
high and their corresponding contributions become negligible. Due to
the fast rise time of the internal PMT of the OB, $\sigma_{OB}$ can
also be neglected. An example of the time difference distribution is
given in Figure~\ref{fig:closeom}. Measurements with several OMs close
to OBs give a time resolution of $\sim$0.5~ns for the electronics, in
agreement with the results obtained during the onshore calibration.

\begin{figure}
 \begin{center}
 \epsfxsize=10.0cm
 \epsffile{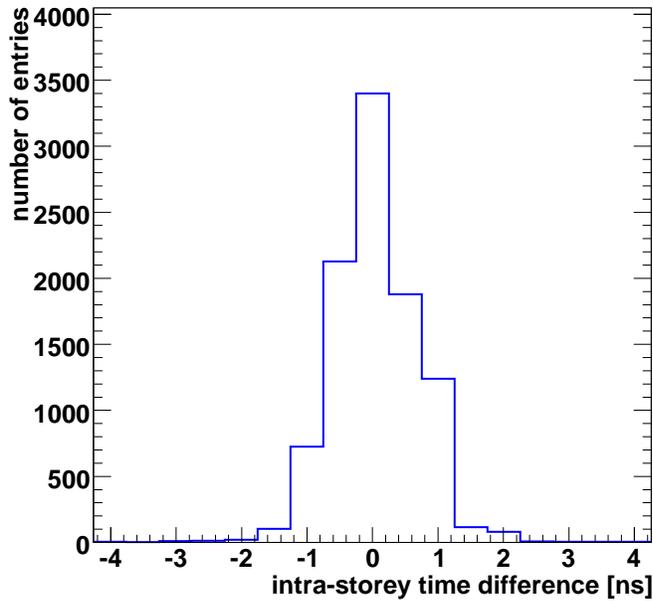}
 \caption{\small Example of the distribution of the hit time
 differences for a pair of OMs in the same storey when illuminated by a
 close OB. The sigma of the distribution (0.7 ns) indicates a
 front-end electronics resolution of 0.7/$\sqrt 2=0.5$~ns.}
 \label{fig:omom}
 \end{center}
\end{figure}

A third way to check this time resolution is given by comparing the
 times measured by two OMs in the same storey when illuminated at high
 intensity. In this case, the sigma of the distribution (see
 Figure~\ref{fig:omom}) is given by $\sigma_{OM-OM}=\sqrt{2}
 \sigma_{elec}$. A front-end electronics time resolution of 0.5~ns is
 obtained.

These checks indicate that the uncertainty introduced by the readout
system is negligible with respect to the irreducible contributions
from the chromatic dispersion and TTS.

\section{Conclusion}
\label{sec:conclusion}

The completion of the ANTARES telescope has opened a new window to the
neutrino Southern sky. Since the main physics goal of ANTARES is the
search for cosmic point sources, an important feature is its angular
resolution (a few tenths of a degree above 10~TeV), unsurpassed for
this kind of instruments. The track reconstruction algorithms are
based on the PDF of the photon arrival times to the PMTs. Therefore,
in order to ensure an optimal performance, a precise time calibration
of the detection system is crucial.

An onshore calibration performed in the laboratory provides a
preliminary time calibration. Once the detector is deployed in the
sea, time calibrations are performed {\it in situ} with a system of
optical beacons. Although most of the corrections to the onshore time
offsets are small, there are 15\% of cases where they are larger than
1~ns. The observation of coincident photons from the decay of $^{40}$K
provides a completely independent verification for these
measurements. 

The adopted calibration systems and methods attain a relative time
calibration between detector elements of less than 1~ns, as required.

\section*{Acknowledgments}

The authors acknowledge the financial support of the funding agencies:
Centre National de la Recherche Scientifique (CNRS), Commissariat
\`{a} l'\'{e}nergie atomique et aux \'{e}nergies alternatives (CEA), Agence
National de la Recherche (ANR), Commission Europ\'{e}enne (FEDER fund
and Marie Curie Program), R\'{e}gion Alsace (contrat CPER), R\'{e}gion
Provence-Alpes-C\^{o}te d'Azur, D\'{e}parte-ment du Var and Ville de
La Seyne-sur-Mer, France; Bundesministerium f\"{u}r Bildung und
Forschung (BMBF), Germany; Istituto Nazionale di Fisica Nucleare
(INFN), Italy; Stichting voor Fundamenteel Onderzoek der Materie
(FOM), Nederlandse organisatie voor Wetenschappelijk Onderzoek (NWO),
the Netherlands; Council of the President of the Russian Federation
for young scientists and leading scientific schools supporting grants,
Russia; National Authority for Scientific Research (ANCS), Romania;
Ministerio de Ciencia e Innovaci\'{o}n (MICINN), Prometeo of Generalitat
Valenciana (GVA) and MultiDark, Spain. We also acknowledge the
technical support of Ifremer, AIM and Foselev Marine for the sea
operation and the CC-IN2P3 for the computing facilities.

\end{document}